\documentclass[aps,prd,eqsecnum,11pt,showpacs,preprintnumbers,nofootinbib]{revtex4-1}
\usepackage{amsmath,amsfonts,amssymb}
\usepackage[pdftex]{graphicx}
\usepackage{bm,color}

\IfFileExists{dsfont.sty}
	{\usepackage{dsfont}
         \let\mathbb=\mathds
         \newcommand{\id}{\mathds{1}}}
	{\typeout{Package dsfont.sty was not found, using alternative macros.}
         \let\mathds=\mathbb
         \newcommand{\id}{\mbox{1 \kern-.59em \textrm{l}}}}

\newcommand{\g}{\gamma}
\renewcommand{\d}{\delta}
\newcommand{\e}{\epsilon}

\newcommand{\ka}{\kappa}
\renewcommand{\l}{\lambda}
\newcommand{\m}{\mu}

\renewcommand{\t}{\tau}

\newcommand{\w}{\omega}
\newcommand{\G}{\Gamma}
\newcommand{\D}{\Delta}

\newcommand{\Th}{\Theta}

\renewcommand{\L}{\Lambda}

\newcommand{\W}{\Omega}
  \newcommand{\cL}{\mathcal{L}}

\newcommand{\bx}{\mathbf{x}}

\newcommand{\be}{\begin{equation}}
\newcommand{\ee}{\end{equation}}
\newcommand{\bes}{\begin{subequations}}
\newcommand{\ees}{\end{subequations}}
\newcommand{\bea}{\begin{eqnarray}}
\newcommand{\eea}{\end{eqnarray}}

\newcommand{\pa}{\partial}

\newcommand{\nn}{\nonumber \\}
\newcommand{\na}{\nabla}

\newcommand{\lag}{\langle}
\newcommand{\rag}{\rangle}

\newcommand{\sdfrac}[2]{\mbox{\small$\displaystyle\frac{#1}{#2}$}}

\def\nbox#1#2{\vcenter{\hrule \hbox{\vrule height#2in
\kern#1in \vrule} \hrule}}
\def\sq{\,\raise.5pt\hbox{$\nbox{.09}{.09}$}\,}
\def\sqb{\,\raise.5pt\hbox{$\overline{\nbox{.09}{.09}}$}\,}
\def\sech{{\rm sech}}

\allowdisplaybreaks

\begin{document}

\preprint{LA-UR-17-28548}
\date{December 12, 2017}

\title{\begin{center}\Large
Decay of the de Sitter Vacuum
\end{center}}

\author{Paul R. Anderson$^{\rm a}$}
\altaffiliation{\tt anderson@wfu.edu}
\author{Emil Mottola$^{\rm b}$}
\altaffiliation{\tt emil@lanl.gov}
\author{Dillon H. Sanders$^{\rm a}$}
\affiliation{\vspace{5mm}$^{\rm a)}$
Department of Physics, \\
Wake Forest University, \\ Winston-Salem,
NC 27109, USA\\}
\affiliation{$^{\rm b)}$ Theoretical Division, T-2, MS B285\\
Los Alamos National Laboratory\\ Los Alamos, NM 87545 USA}

%\pacs{\ 04.62.+v,\ 95.36.+x,\ 98.80.Qc}

\begin{abstract}
\vspace{-2mm}\noindent

The decay rate of the Bunch-Davies state of a massive scalar field in the expanding flat spatial sections
of de Sitter space is determined by an analysis of the particle pair creation process in real time. The  
Feynman definition of particle and antiparticle Fourier mode solutions of the scalar wave equation, 
and their adiabatic phase analytically continued to the complexified time domain, show conclusively that the
Bunch-Davies state is not the vacuum state at late times. The closely analogous creation of charged 
particle pairs in a uniform electric field is reviewed and Schwinger's result for the vacuum decay rate 
is recovered by the real time analysis. The vacuum decay rate in each case is also calculated by switching 
the background field on adiabatically, allowing it to act for a very long time, and then adiabatically switching 
it off again. In both the uniform electric field and de Sitter cases the particles created while the field is switched 
on are verified to be real, in the sense that they persist in the final asymptotic flat zero-field region. In the de Sitter
case there is an interesting residual dependence of the rate on how the de Sitter phase is ended, indicating
a greater sensitivity to spatial boundary conditions. The electric current of the created particles in the $E$-field case 
and their energy density and pressure in the de Sitter case are also computed, and the magnitude of their backreaction 
effects on the background field estimated. Possible consequences of the Hubble scale instability of the de Sitter vacuum 
for cosmology, vacuum dark energy, and the cosmological `constant' problem are discussed.
\vspace{-5cm}

\end{abstract}
\maketitle

\setcounter{page}{1}

\section{Introduction}
\label{Sec:Intro}

The vacuum state of quantum field theory (QFT) in flat Minkowski space, with no external fields,
is defined as the eigenstate of the Hamiltonian with the lowest eigenvalue. The existence of a Hamiltonian
generator of time translational symmetry, with a non-negative spectrum, bounded from below is
crucial to the existence and determination of the vacuum ground state, containing no particle excitations.
Particle states are defined then by solutions of relativistic wave equations forming irreducible representations of the
Poincar\'e group. The vacuum state is invariant under translations, rotations and Lorentz boosts, and the 
correlation functions built upon this vacuum state enjoy complete invariance under Poincar\'e symmetry.

As is well known, these properties do not hold in a general curved spacetime, in time dependent external background fields, 
nor even for a free QFT in flat spacetime under general coordinate transformations that are not Poincar\'e symmetries. In 
these cases the Hamiltonian becomes time dependent or no Hamiltonian bounded from below exists at all, and the concepts 
of `vacuum' or `particles'  become much more subtle. In situations when the background has a high degree of symmetry, such 
as de Sitter spacetime, it has been customary to avoid the particle concept altogether, and focus attention instead on the state 
possessing the maximal symmetry of the background. In de Sitter space this maximal $O(4,1)$ symmetric state for massive fields 
is commonly known as the Bunch-Davies state.\footnote{The state was first investigated by several authors \cite{CherTag,BunDav} 
and might also be called the Chernikov-Tagirov-Bunch-Davies (CTBD) state.} The question of whether the Bunch-Davies state is 
actually a `vacuum' state, or a stable state at all has been the subject of a number of investigations \cite{MotPRD85,Poly,AndMot1,AndMot2}, 
although with implications that appear to differ somewhat from each other, even for free fields \cite{DolEinZak,Attract}. When 
self-interactions are considered, additional differences between the various approaches arise
\cite{Myhr,AkhBui,Hig,BroEpMos,MarMor,Ser,BoyHol,JatLebRaj,Akh,Hol,AkhMosPavPop}. At yet another level are the potential effects 
of graviton loops, when higher order gravitational interactions are considered \cite{Wood}.

In view of the central role the Bunch-Davies state plays in cosmological models of inflation and the origin of fluctuations that give
rise to anisotropies in the universe \cite{DodMuk}, as well as the importance of de Sitter vacuum instability to the fundamental 
issue of vacuum energy in cosmology and the cosmological constant problem \cite{Wein,Pad,NJP}, it is essential that the physical 
basis of the QFT vacuum be clearly established. Reconciling the various approaches to vacuum energy in cosmology 
when the technical issues that arise in the cases interactions, light fields, or graviton loops are considered, is bound to be more difficult 
if de Sitter vacuum decay in the simplest and best controlled case of a massive scalar free field is not first fully clarified.

To this end in this paper we discuss in detail the close correspondence between QFT in de Sitter spacetime and in the non-gravitational 
background of a constant, uniform electric field \cite{AndMot1}, which provides important guidance for the de Sitter case. Both backgrounds 
have a high degree of symmetry, which permit exact solutions and natural generalizations of concepts and QFT methods 
from the case of a flat, zero-field background. Yet neither admit a conserved Hamiltonian bounded from below, and in
both cases particle creation occurs and vacuum decay is expected.

In the case of a constant, uniform electric field, the QFT of charged matter, neglecting self-interactions, was considered by Schwinger 
in the covariant proper time representation \cite{Schw}. By this covariant heat kernel method the vacuum decay probability and decay rate 
in terms of the imaginary part of the one-loop effective action, defined by analytic continuation in the proper time variable is obtained. 
Since the $E$-field spontaneously decays into particle/antiparticle pairs, the `vacuum' is {\it not} the state of maximal symmetry of the 
background (which is time reversal symmetric), but instead the $E$-field initiates a non-trivial time dependent process, which almost 
certainly leads to a state populated with particle/antiparticle excitations in which the coherent mean electric field vanishes asymptotically 
at late times. The correspondence with the de Sitter case suggests that cosmological vacuum energy should similarly decay into particle/antiparticle pairs, 
eventually leading to a state with small but non-zero slowly decaying vacuum energy \cite{MotPRD85,NJP}.

Schwinger's proper time method makes no explicit reference to particles, and its very elegance disguises somewhat the physical  
definition of vacuum it entails. Later studies of QFT in a constant $E$-field by canonical quantization methods \cite{Nar,Nik,NarNik,FradGitShv,GavGit}
revealed that the essential ingredient is the $m^2 \rightarrow m^2 - i \e$ prescription where $\e \rightarrow 0^+$. This is of course the same $i \e$ 
prescription defining the causal propagator function in flat space QFT, which Feynman obtained by identifying positive frequency solutions of the 
wave equation as particles propagating forward in time, and negative frequency solutions as the corresponding antiparticles propagating backward 
in time \cite{Feyn}. It is this physical condition that provides the mathematically precise definition of particle/antiparticle excitations and fixes the 
vacuum state of relativistic QFT in {\it real} time, which by continuity and the adiabatic theorem applies also to background fields or weakly curved spacetimes.

The $m^2 - i \e$ rule is specified in the Schwinger-DeWitt approach by a single exponential describing a relativistic particle worldline. 
Thus causality is enforced by the particle always moving forward in its own proper time, whether the external coordinate time does so or not. This observation,
first made by Stueckelberg \cite{Stuk}, carries over unaltered to curved spacetimes. The extension of the covariant Schwinger heat kernel method 
to gravitational backgrounds was developed by DeWitt \cite{DeWitt}. The causality condition and $m^2$ analyticity it implies in the covariant 
Schwinger-DeWitt formulation was later shown to be equivalent to the requirement that ingoing particle modes as $t \rightarrow -\infty$ are analytic in 
the {\it upper} half complex $m^2$ plane, while the corresponding outgoing modes as $t \rightarrow +\infty$ are analytic in the {\it lower} half complex 
$m^2$ plane \cite{Rumpf,Rumpf81}. Antiparticle modes are the complex conjugate solutions of the wave equation in which the analyticity requirements 
in the upper/lower mass  squared plane are reversed. It is not difficult to see that the definition of the corresponding $|0, in \rag$ and $|0, out \rag$ vacuum 
states, and the {\it in-out} effective action they imply, leads to a non-trivial Bogoliubov transformation between the $|0,in\rag$ and $|0, out \rag$ states, 
particle pair creation and a vacuum decay rate for de Sitter space analogous to that of the uniform $E$-field \cite{MotPRD85,AndMot1}.

In the canonical description in terms of Fock space creation and destruction operators, pair creation manifests itself
as a non-trivial Bogoliubov transformation between the positive/negative frequency operators as $t\rightarrow -\infty$,
which define the $|0, in\rag$ vacuum, relative to the corresponding positive/negative frequency operators as $t\rightarrow +\infty$
which define the $|0, out\rag$ vacuum. The overlap probability
\vspace{-3mm}
\be
\big\vert \lag 0, out |0, in\rag\big\vert^2_{_{V,T}} = \exp\big\{\!-VT\, \G\big\} = \exp\big\{\!-2\,VT \,{\rm Im} \cL_{\rm eff}\big\}
\label{inoutrate}
\ee
behaves exponentially in the spatial volume $V$ and time $T$ that the background field is applied, and $\G= 2\, {\rm Im}\, \cL_{\rm eff}$
is twice the imaginary part of the effective Lagrangian density $\cL_{\rm eff}$  found by Schwinger in the proper time approach. To be 
meaningful the four-volume factor ${\cal V}_4 = VT$ must be removed from (\ref{inoutrate}), with $\cL_{\rm eff}$ independent of the spacetime
coordinates extracted, for backgrounds independent of $\bf x$ and $t$. Then $\G$ is the constant decay rate of the $|0, in\rag$ vacuum per unit 
volume per unit time due to steady spontaneous creation of particle/antiparticle pairs by the fixed classical ($E$ or de Sitter) background field, 
under the assumption that the space and time dependence of the background and any backreaction may be neglected at lowest order. 
Notice that the effective action and the vacuum decay rate given by its imaginary part in (\ref{inoutrate}) are coordinate invariant quantities, even though 
the space+time splitting, and definition of positive/negative frequency modes, Bogoliubov transformation, and definition of particles is not. 
General coordinate invariance is manifest throughout only in the worldline proper time representation.

It is important also to realize that the imaginary part of the one-loop effective Lagrangian ${\rm Im}\, {\cal L}_{\rm eff}$, and vacuum decay rate $\G$ 
cannot be obtained by reliance upon a calculation in Euclidean time $\t = it$, but instead requires definition of the vacuum consistent
with causality in {\it real} time. Indeed, since the electric field is the $F_{j0}$ component of the field strength tensor, a Euclidean calculation with 
$E_j= F_{j0} \rightarrow i F_{j4}$ in this case would be tantamount to treating it as a constant {\it magnetic} field background $B_i = \frac{1}{2}\e_{ijk} F_{jk}$, 
for which the quantum Hamiltonian is bounded from below, particle trajectories are circular rather than hyperbolic, no particle creation at all occurs, and 
the vacuum is stable, all of which is completely different physics than the $E$-field background in real time. Including particle self-interactions does not 
change these fundamental differences between backgrounds for which the Lorentz invariant ${\bf B}^2- {\bf E}^2$ have opposite signs. An interacting QFT 
built upon the $B$-field vacuum with Euclidean time correlation functions is therefore necessarily {\it physically inequivalent} to the $E$-field in real time, 
completely missing the particle pair creation and vacuum decay rate contained in (\ref{inoutrate}), {\it even at lowest zeroth order} in the self-interactions. 

This essential difference between specification of the vacuum in real time and the postulate of Euclidean analyticity is one root cause of some of 
the different results and claims in the literature. The difference between the $m^2 - i \e$ prescription {\it vs.} Euclidean continuation is not simply a 
difference of formalisms, but rather of enforcing completely different physical requirements on the QFT vacuum by different initial/final conditions 
in real time than those imposed by regularity in the Euclidean time domain. Although equivalent in flat Minkowski space, the $m^2 - i \e$ prescription 
required by causality is mathematically {\it inconsistent} with Euclidean continuation in background fields such as the $E$-field or de Sitter space, 
and leads to physically different results.

One way of seeing why the equivalence of Euclidean continuation to the causal QFT vacuum in flat space no longer holds in de Sitter space is in the 
qualitatively different properties of representations of the Poincar\'e and de Sitter symmetry groups. It is an important special property of Minkowski 
spacetime that the subspaces of positive and negative frequency solutions of the wave equation are {\it separately} invariant under proper orthochronous 
Lorentz transformations, so that the Stueckelberg-Feynman definition of vacuum is consistent with maximal Poincar\'e symmetry. In contrast, there is 
{\it no} $SO(4,1)$ invariant decomposition of positive and negative frequency solutions in de Sitter spacetime. Any such decomposition into positive 
and negative frequency subspaces mix under $SO(4,1)$ symmetry transformations, and transform with equivalent representations of the de Sitter 
group~\cite{Nacht}. As a result there is no de Sitter invariant way to distinguish particles from antiparticles, and no reason for the physical 
Stuekelberg-Feynman definition of particle excitations or vacuum to lead to a de Sitter invariant state. Indeed on any finite time slice, it does not. 

Closely related to and following from analyticity requirements in $m^2$, rather than a Euclidean postulate, is the fact that the de Sitter/Feynman
propagator $G_F(x, x')$ calculated with $in$-$out$ boundary conditions obeys the composition rule
\be
\int_{\Sigma_x} d \Sigma^{\m}_x\  G_F(x_1, x)\stackrel{\leftrightarrow}{\na}_{\m}G_F(x, x_2) = G_F(x_1, x_2)
\label{comprule}
\ee
consistent with causality, where $\Sigma_x$ is an arbitrary spacelike surface intermediate between $x_1$ and $x_2$. 
This composition rule again expresses the Stueckelberg-Feynman prescription of particles moving forward in time, antiparticles 
backward in time, and results from the representation of $G_F$ in terms of a single exponential in worldline 
proper time in Schwinger's method, rather than a sum of exponentials of opposite sign. These single exponentials combine
simply and lead to the composition (\ref{comprule}), which is {\it violated} by the Euclidean Bunch-Davies propagator \cite{Poly}.
In physical terms the Bunch-Davies state is best understood not as a `vacuum' state at all, but as a particular finely tuned phase
coherent superposition of particle and anti-particle modes \cite{AndMot1}.

In this paper we study the particle creation process in greater detail in real time, in the flat Poincar\'e spatial slicing of de Sitter
spacetime most commonly considered in cosmology, and compute the decay rate of the Bunch-Davies `vacuum' state by two different
methods. The particle creation process is studied in real time by analysis of the adiabatic phase integral and its analytic
extension to the complex time domain. This analysis allows for determination of which Fourier modes are experiencing
creation `events' in each interval of time that the external de Sitter or $E$-field background is applied, removing the infinite
four-volume factor ${\cal V}_4$ and determining the constant finite rate $\G$ and pre-factor in (\ref{DecayRateDiff}) unambiguously.

We also consider the vacuum decay rate $\G$ obtained by adiabatically turning the electric or expanding de Sitter background 
field on and then off again, after the lapse of a long but finite time $T$. In this approach there can be no ambiguity of initial and final
vacuum states, since the geometry at both early ($t \rightarrow -\infty$) and late ($t \rightarrow +\infty$)
times is Minkowski flat space with zero field. This approach also eliminates the somewhat unsatisfactory feature of 
previous constant $E$-field or de Sitter background calculations, in that a formal divergence in the sum over modes appearing
in those calculations, which must be cut off by appeal to the finite `window' of modes undergoing particle creation in a finite time
$T$ in the constant background, is explicitly removed and regulated and a finite $\G$ obtained. This second method also reveals 
the sensitivity of de Sitter space to correlations over superhorizon scales that may have important implications for cosmology.

The outline of the paper is as follows. The next section reviews the general framework of particle creation and vacuum decay
as a violation of the adiabatic condition in persistent background fields. Sec. \ref{Sec:PersistE} shows how the adiabatic phase
integral, and its Stokes lines of constant Real Part in the complex plane can be used  to determine the time of particle creation
events, which applied to the case of a constant, uniform electric field reproduces Schwinger's result for the vacuum decay of the
$E$-field. In Sec. \ref{Sec:Eadb} two time profiles of the uniform electric field for which it is adiabatically turned on and off with a
long duration $T$ in between where it is constant are used to compute the particle creation and decay rate in the limit $T \rightarrow \infty$,
again reproducing the Schwinger result. The electric current is also computed for one profile and shown to grow linearly with $T$,
so that the secular effects of backreaction clearly must be taken into account for persistent fields. In Sec. \ref{Sec:dSPersist}, the
same adiabatic phase method is applied to a persistent de Sitter background in the spatially flat Poincar\'e coordinates, and the
particle creation and finite decay rate of the Bunch-Davies state determined. In Sec. \ref{Sec:dSadb} we consider two time profiles
for which  the de Sitter background is adiabatically turned on and off with a long time duration in between, similar to the $E$-field case.
For one profile we evaluate the particle creation and decay rate numerically in the limit of long time duration. In Sec. \ref{Sec:Ener} an 
analytic estimate is made of the energy density and pressure of the particles created in the de Sitter phase along with an estimate of 
the strength of their backreaction effect. Sec. \ref{Sec:Concl} contains a discussion of our conclusions including the possible implications 
for inflation, vacuum dark energy, and the cosmological `constant' problem.

\section{Particle Creation and the Adiabatic Phase}
\label{Sec:Phase}

We consider a quantum field interacting only with a classically prescribed ({\it i.e.} non-dynamical) external background field.
For simplicity we specialize to a non-self-interacting scalar field $\Phi$ and a spatially homogeneous but time-dependent classical
background. Making use of spatial homogeneity, the quantum scalar field operator in flat space may be expanded in a Fourier series
\be
\Phi (t, {\bf x}) = \frac{1}{\sqrt{V}\ } \sum_{\bf k} \left\{a_{\bf k}\, e^{i{\bf k \cdot x}}\, f_{\bf k}(t) +
b_{\bf k}^{\dag}\, e^{-i{\bf k \cdot x}}\, f_{\bf k}^*(t) \right\}
\label{Phiop}
\ee
whose time-dependent mode functions obey second order differential equations in time of the form
\be
\left[ \frac{d^2}{dt^2} + \w_{\bf k}^2(t) \right] f_{\bf k}(t) = 0 \;.
\label{harmosc}
\ee
Here $\bf k$ labels the spatial momentum which takes on discrete values for periodic boundary conditions in the finite
volume $V$. For the cases of interest in this paper $\w_{\bf k}^2(t)$ is a real time-dependent frequency function that 
for massive fields is strictly positive, nowhere vanishing on the real time axis $-\infty < t< \infty$.

Considering the case of a charged complex scalar field $\Phi (t, {\bf x})$ in a pure electric
field background ${\bf E}$, the Klein-Gordon wave equation $(\pa_\m - i e A_\m)^2\Phi = m^2 \Phi$ gives
\be
\w_{\bf k}^2(t) = \big({\bf k} - e {\bf A}(t)\big)^2 + m^2
\label{freq}
\ee
in the $A_0 =0$ gauge in which ${\bf E} = - \dot {\bf A}$. The case of an uncharged Hermitian scalar field (obeying $\Phi^{\dag} = \Phi$
and $a_{\bf k} = b_{\bf k}$) in a spatially homogeneous and isotropic cosmological spacetime may also be reduced to
a mode equation of the form (\ref{harmosc}) with a different $\w_k(t)$: {\it cf.} (\ref{FLRWomega}).

The complex valued solutions of (\ref{harmosc}) are required to satisfy the Wronskian condition
\be
f_{\bf k} \dot f_{\bf k}^* - \dot f_{\bf k} f_{\bf k}^* = i \hbar
\label{Wron}
\ee
constant in time, and the Fock space operators are required to obey the commutation relations
\be
[a_{\bf k}, a^{\dag}\,_{\!\!\bf k'}] = [b_{\bf k}, b^{\dag}\,_{\!\!\bf k'}] = \d_{{\bf k}, {\bf k'}}
\ee
in order for the Heisenberg field operator $\Phi$ to satisfy the canonical equal time commutation relation
\be
\Big[\Phi (t, {\bf x}), \frac{\partial \Phi^{\dag}}{\partial t} (t, {\bf x'}) \Big] = i \hbar\, \delta^3({\bf x} - {\bf x'})
\ee
in a finite volume $V$. In the absence of any external field, ${\bf A}$ and $\w_{\bf k}$ are constants, and
\be
f_{\bf k}(t) = f_{\bf k}^{(0)}(t)\equiv \sqrt{\sdfrac{\hbar}{2 \w_{\bf k}}} \, e^{- i \w_{\bf k}t}
\label{fkflat}
\ee
defines the {\it positive} energy particle mode function that is analytic in the {\it upper} half complex $m^2$ plane in both
limits $t \rightarrow \pm \infty$. The corresponding Minkowski no-particle state $| 0\rag$ defined by
\be
a_{\bf k}\, | 0 \rag = b_{\bf k}\, | 0\rag = 0\,, \qquad \forall\quad{\bf k}
\label{flatvac}
\ee
is both the vacuum $|0, in \rag$ state and the vacuum $|0, out \rag$ state for all times, and there is no
spontaneous particle creation or vacuum instability in this case.

The physical basis for extending the definition of no-particle vacuum states to the case of slowly varying weak external
fields is based upon the adiabatic theorem, which guarantees that the state of a quantum system does not change
if subjected to an external perturbation that is arbitrarily slowly varying in time \cite{BornFock}. Hence in weak or slowly
varying external fields the QFT vacuum must be `close' to that (\ref{fkflat}) and well-determined up to small terms in an 
asymptotic expansion of the solution of (\ref{harmosc}) in terms of the time derivatives of $\w_{\bf k}$. The adiabatic phase integral
\be
\Th_{\bf k}(t) = \int^t dt' \, \w_{\bf k}(t')
\label{genTheta}
\ee
then takes on fundamental importance, since the zeroth order adiabatic mode function
\be
\tilde f^{(0)}_{\bf k}(t) \equiv \frac{1}{\sqrt{2\w_{\bf k}(t)}}\, \exp \big\{\!- i\, \Th_{\bf k}(t)\big\}
\label{adbzero}
\ee
is an approximate positive frequency (particle) solution to (\ref{harmosc}) satisfying (\ref{Wron}) (with $\hbar = 1$ hereafter)
in the limit that $\w^2_{\bf k}(t)$ is a slowly varying function of $t$. Higher order adiabatic mode functions $\tilde f^{(n)}_{\bf k}$
may be found by substituting the exponential ansatz
\be
f_{\bf k}(t) = \frac{1}{\sqrt{2W_{\bf k}(t)}}\, \exp \left\{ - i\, \int^t dt ' W_{\bf k}(t')\right\}
\ee
into the mode eq. (\ref{harmosc}), resulting in the {\it exact} nonlinear equation for $W_{\bf k}(t)$
\be
W_{\bf k}^2 = \w_{\bf k}^2 + \frac{3}{4} \frac{\dot W_{\bf k}^2}{ W_{\bf k}^2} - \frac{1}{2} \frac{\ddot W_{\bf k}}{W_{\bf k}}
\ee
which may then be expanded in an asymptotic series in time derivatives:
\be
W_{\bf k} = \w_{\bf k} \left\{ 1 + \frac{3}{8} \frac{\dot \w_{\bf k}^2}{ \w_{\bf k}^4} - \frac{1}{4} \frac{\ddot \w_{\bf k}}{\w_{\bf k}^3} + \dots \right\}\,.
\label{asymexp}
\ee
Clearly the lowest (zeroth) order adiabatic mode function (\ref{adbzero}) with $W_{\bf k}^{(0)}(t) = \w_{\bf k}(t)$ is a good approximation
to the solution of (\ref{harmosc}) and the adiabatic theorem is applicable only to the extent that the relative size of the corrections
in (\ref{asymexp}) parametrized by
\be
\vert \d_{\bf k}(t)\vert  \equiv \left\vert\frac{3}{8} \frac{\dot \omega_{\bf k}^2}{\omega_{\bf k}^4}
- \frac{1}{4} \frac{\ddot \omega_{\bf k}}{\omega_{\bf k}^3}\right\vert \ll 1
\label{adbcond}
\ee
remain uniformly small for all $t$. 

For $\bf k$ such that (\ref{adbcond}) holds, $\tilde f^{(0)}_{\bf k}(t)$ remains an approximate positive frequency
particle solution with the required analyticity in $m^2$, and particle creation is negligibly small. Since $\vert\d_{\bf k}\vert \rightarrow 0$
as $|\bf k| \rightarrow \infty$, the adiabatic condition (\ref{adbcond}) does hold arbitrarily accurately in this limit for smoothly varying background
fields with bounded time derivatives. Thus there is no particle creation in arbitrarily high momentum modes, and the vacuum remains the
vacuum at large momenta or short distances. It is just this property that makes the adiabatic expansion useful for renormalization of composite
operators such as the electric current or energy-momentum tensor in smoothly time varying backgrounds, requiring only the standard
counterterms expected on the basis of usual power counting arguments \cite{BirBun,BirDav}.

If on the other hand the condition (\ref{adbcond}) fails to hold at some times, and particularly at small to moderate $|\bf k|$, on the scale
of the time variation of the background field, these Fourier modes will then receive some admixture of the complex conjugate
approximate solution to (\ref{adbzero}). Because of the association of the complex conjugate solution to negative energy
or antiparticle modes in (\ref{Phiop}) by the $m^2 - i\e$ prescription, the violation of the adiabatic approximation near the maxima
of $\vert \d_{\bf k}(t)\vert$ corresponds to particle/antiparticle pairs being created spontaneously from the vacuum \cite{Parker,Zel,ParFul}.

The adiabatic mode functions (\ref{adbzero}), perhaps extended by use of a frequency function of higher order in the asymptotic
expansion (\ref{asymexp}), also provide useful templates against which the exact mode function solutions of (\ref{harmosc}) may
be compared. The transformation between the two bases of $f_{\bf k}$ and $\tilde f_{\bf k}^{(n)}$ defines a time-dependent Bogoliubov
transformation which may be used to define a semi-classical time-dependent particle number  \cite{QVlas,HabMolMot,ShortDistDecohere,AndMot1}
\vspace{-1mm}
\be
{\cal N}_{\bf k}^{(n)}(t) = \frac{1}{2W_{\bf k}^{(n)}\!}\ \bigg\vert\dot f_{\bf k} + \Big(iW_{\bf k}^{(n)} - \sdfrac{V_{\bf k}^{(n)}\!\!}{2\,}\,\Big)\,f_{\bf k}\bigg\vert^{\,2}
\label{adbpart}
\ee
with respect to the $n^{th}$ order adiabatic basis functions defined by the pair of real time-dependent functions $W_{\bf k}^{(n)}(t), V_{\bf k}^{(n)}(t)$
chosen to match the asymptotic expansion (\ref{asymexp}) for $W_{\bf k}$ and $-\dot W_{\bf k}/W_{\bf k}$ respectively to $n^{th}$ order in
time derivatives. This definition is {\it local in time}, and has some necessary arbitrariness in that a choice of adiabatic order $n$ must be made.
Generally the lowest order $n=0,1,2$ approximations are the most useful for applications, such as defining an approximate particle number
density for transition to a semi-classical Boltzmann-Vlasov transport description of non-equilibrium relativistic quantum systems \cite{QVlas},
and in particular allowing dissipative particle interactions to be taken into account.

In this paper we focus on the application of the adiabatic method to background fields that are persistent for long periods of time,
such as constant uniform electric fields, or de Sitter space, where in both cases the adiabatic condition (\ref{adbcond}) holds arbitrarily
accurately {\it asymptotically} as $t \rightarrow \mp \infty$, for any {\it finite} $\bf k$. In these cases asymptotic $|0,in\rag$ and $|0, out \rag$
vacuum states can be defined, accompanied by well-defined particle number Fock basis operators. Since the condition (\ref{adbcond})
is not satisfied at all intermediate times for some $\bf k$, the approximate solution (\ref{adbzero}) is not a uniformly valid solution to
the exact eq.~(\ref{harmosc}) for all $t$, and particle creation occurs in such non-trivial persistent background fields. The specification
of vacuum states in the asymptotic past or future is necessarily a {\it global in time} definition, that describes {\it secular} or long time effects.

In this case the task is to determine the admixture of the negative frequency complex conjugate solution at late times,
$t \rightarrow +\infty$, given that the exact solution $f_{\bf k}(t)$ is a pure positive frequency solution of the form (\ref{adbzero})
at early times, $t\rightarrow -\infty$, or in other words, to determine the {\it time-independent} Bogoliubov coefficients $(A_{\bf k}, B_{\bf k})$
for the exact solutions of (\ref{harmosc}) satisfying the asymptotic conditions
\vspace{-3mm}
\be
f_{\bf k} (t) \rightarrow \left\{\begin{array}{ll} \tilde f^{(0)}_{\bf k}(t)\,, \quad & t \rightarrow -\infty\\
A_{\bf k}\, \tilde f^{(0)}_{\bf k}(t) + B_{\bf k}\, \tilde f^{(0)*}_{\bf k}(t)\,, \quad &  t \rightarrow + \infty
\end{array}\right.
\label{asymBog}
\ee
which has the form of a one-dimensional scattering problem. It is {\it `over-the-barrier'} scattering if $\w_{\bf k}^2(t)$ is strictly
positive for all real $t$, so that there are no classical turning points on the real $t$ axis.

Because of the Wronskian condition (\ref{Wron}), the Bogoliubov coefficients necessarily satisfy
\be
\vert A_{\bf k}\vert^2 - \vert B_{\bf k}\vert^2 = 1
\label{Bogcond}
\ee
characteristic of a time-independent canonical transformation. Because of this condition, the Bogoliubov coefficients may be characterized
by a hyperbolic angle parameter $\chi_{\bf k}$. In the second quantized description (\ref{Phiop}) the quantity $\vert B_{\bf k}\vert^2
= \sinh^2 \chi_{\bf k}$ is the well-defined mean number density of particles at asymptotically late times created in the mode $\bf k$ by the background
electric or gravitational field. The coefficient $\vert B_{\bf k} \vert^2$ may be calculated in some special cases such as the constant $E$-field
and de Sitter backgrounds by knowledge of the exact scattering solutions of (\ref{harmosc}) satisfying (\ref{asymBog}), or approximately
by the complex WKB adiabatic phase methods to be discussed in the next section, or finally, by direct numerical solution of the mode eq.~(\ref{harmosc}).

Since the vacuum state for a non-self-interacting field theory is a product of Gaussian harmonic oscillator wave functions, one for each
$\bf k$, it is a straightforward exercise to represent the initial Gaussian state and Fock space operators in the final state basis, in terms
of $\vert B_{\bf k} \vert^2$. For a single real hermitian scalar field beginning in the $|0,in\rag$ vacuum, the diagonal elements of the
Gaussian density matrix $\varrho$ for the $n^{th}$ excited state of the oscillator labelled by $\bf k$ in the final $|n,out\rag$ state basis are \cite{BroCar,CKHM}
\be
\varrho_{n}({\bf k})\big\vert_{{\rm Real}\,\bf \Phi} = |\lag n, out \vert 0, in\rag|^2 = \d_{n, 2 \ell}\ \frac{(2\ell)!}{4^\ell(\ell !)^2}
\ \sech\,\chi_{\bf k} \, (\tanh \chi_{\bf k})^{2 \ell}
\label{rhopair}
\ee
which in the second quantized Fock space description (\ref{Phiop}) is the probability of finding $n=2 \ell$ particles in the Fourier mode $\bf k$
in the final state, if none were present in the initial state. In (\ref{rhopair})
\be
\tanh^2 \chi_{\bf k} = \frac{|B_{\bf k}|^2}{|A_{\bf k}|^2}\,,\qquad \sech\,\chi_{\bf k} = \frac{1}{|A_{\bf k}|} = \left[ 1 + |B|^2_{\bf k}\right]^{-\frac{1}{2}}
\label{tansec}
\ee
with the vanishing of $\varrho_n$ for $n$ odd the result of the fact that the particles can only be created in pairs. Thus the probability that no
particle pairs at all are produced in any mode,
\be
\vert \lag 0,out |0,in\rag\vert^2\big\vert_{{\rm Real}\,\bf \Phi}  = \prod_{\bf k} \varrho_0({\bf k})=  \prod_{\bf k} \sech\,\chi_{\bf k} =
\exp\bigg\{- \sdfrac{1}{2}\sum_{\bf k}\, \ln \Big(1 + |B_{\bf k}|^2\Big)\bigg\}
\label{inout}
\ee
is the vacuum persistence probability, or the probability that the $|0, in \rag$ vacuum at early times will be found in the $|0, out\rag$ vacuum
at late times. Eq. (\ref{inout}) relates the probability of vacuum decay directly to particle creation via the number density of created particles
$|B_{\bf k}|^2$ in the final state defined by the one-dimensional scattering problem (\ref{asymBog}) for spatially homogeneous background fields,
in the {\it in-out} formalism of QFT, enforcing the Feynman-Schwinger $m^2-i\e$ prescription.

For a complex charged scalar field the corresponding diagonal elements of the density matrix are more simply given by \cite{QVlas}
\vspace{-1mm}
\be
\varrho_{2\ell}({\bf k})\big\vert_{{\rm Complex}\,\bf \Phi} = |\lag n, out \vert 0, in\rag|^2 = \d_{n, 2 \ell}\ (\sech\,\chi_{\bf k})^2 \, (\tanh \chi_{\bf k})^{2 \ell}
\label{rhopairch}
\ee
with (\ref{tansec}) as before, so that the corresponding vacuum persistence probability is
\be
\vert \lag 0,out |0,in\rag\vert^2\big\vert_{{\rm Complex}\,\bf \Phi}  = \prod_{\bf k} \varrho_0({\bf k})=  \prod_{\bf k} \sech^2\,\chi_{\bf k} =
\exp\bigg\{-\sum_{\bf k}\, \ln \Big(1 + |B_{\bf k}|^2\Big)\bigg\}
\label{inoutcharg}
\ee
for a charged scalar field decaying into pairs \cite{Nar,Nik,NarNik,FradGitShv}. The relative factor of $2$ between the exponents
of (\ref{inout}) and (\ref{inoutcharg}) may be understood as a result of the doubling of degrees of freedom and the one-loop effective action
for a complex field relative to a real one. In each case one may check from (\ref{rhopair}) and (\ref{rhopairch}) that
$\sum_{\ell =0}^{\infty} \rho_{2\ell}= 1$ so that probability (unitarity) is conserved.

If the external field producing the particles persists over an infinitely long time homogeneously over an infinite volume, $|B_{\bf k}|^2$ becomes independent
of some components of $\bf k$, and the sum  over $\bf k$ in (\ref{inout}) or (\ref{inoutcharg}) diverges. This divergence is not a pathology, but simply a
consequence of a persistent spatially homogeneous external field producing particles at a finite rate everywhere in space for an infinite time. The spatial
volume factor is easily extracted by the usual method of replacing the sum $\sum_{\bf k}$ over discrete Fourier modes by the continuous Fourier integral
$V\int \frac{d^3 {\bf k}}{(2 \pi)^3}$, and dividing the exponent in (\ref{inout})-(\ref{inoutcharg}) by $V$ in the infinite volume continuum limit $V \rightarrow \infty$.
The vacuum decay rate $\G$ per unit volume might be defined then by an expression of the form (\ref{inout}) or (\ref{inoutcharg}), with
\be
\G = \lim_{T \rightarrow \infty} \lim_{V \rightarrow \infty} \frac{1}{VT} \frac{N}{2}
\sum_{\bf k}\, \ln \big(1 + |B_{\bf k}|^2\big) = \frac{N}{2}\lim_{T \rightarrow \infty}
\frac{1}{T}\int \frac{d^3 {\bf k}}{(2 \pi)^3}\, \ln \big(1 + |B_{\bf k}|^2\big)
\label{DecayRateInt}
\ee
where $N=1, 2$ refers to the number of independent scalar fields undergoing particle creation, one for a single real
field, two for a complex charged scalar. As has been remarked previously \cite{Nar,Nik,NarNik,AndMot1}, the integral in (\ref{DecayRateInt})
still diverges for background external fields that persist for an infinite interval of time, and the expression is indeterminate. In order to compute
a well-defined rate of vacuum decay, it is necessary either to turn on the external background field only for a {\it finite} time, or alternately,
to analyze the particle creation process in real time, to determine the finite subset or `window' of Fourier $\bf k$  modes that experience
particle pair creation during specific intervals of time.

In the first method---to be called the Integral (I) Method---one replaces the persistent background field of interest, such as de Sitter space
which extends infinitely far into the past and future, and which is responsible for the divergent $\bf k$ integral, by a substitute external
background field which is turned on slowly around some initial time $t_0$, persists for a very long but finite time $T$, and then is turned off
again slowly at a later time $t_1$. Since for any fixed finite $T$ only a finite range of Fourier modes will undergo particle creation events,
$|B_{\bf k}^2|$ will vanish rapidly outside of a finite window in Fourier space and the integral over $\bf k$ in (\ref{DecayRateInt}) will be finite,
but proportional to $T$. Then one can divide by $T$ and explicitly take the $T\rightarrow \infty$ limit indicated in (\ref{DecayRateInt})
to obtain a finite result for the vacuum decay rate.

This Integral Method has the advantage of defining zero field regions in the infinite past ($t \ll t_0$) and infinite future ($t \gg t_1$) where
particles and vacuum states are unambiguously defined by the standard flat Minkowski space prescription (\ref{fkflat})-(\ref{flatvac}). However
for this method to work, it is essential that a suitable substitute time-dependent external field be found for which the turning on and off of the
background of interest around $t_0$ and $t_1$ be gentle enough not itself to create significant numbers of particles by violation of the adiabatic
condition (\ref{adbcond}), {\it and} for which  any `edge effects' of particle creation around $t_0$ and $t_1$ become negligible in the long time limit
$T \rightarrow \infty$. An example of an external field time profile satisfying these criteria and application of the Integral Method to the uniform
$E$-field case is provided by (\ref{Eprofile}), shown in Fig. \ref{Fig:Eprofile}, with the result of Sec. \ref{Sec:Eadb} for the vacuum decay rate
agreeing with Schwinger's result in the limit $T \rightarrow \infty$. In this flat space example $VT$ is simply the total four-volume
$\int d^4 x = {\cal V}_4$ over which the external $E$-field acts, and $\G = 2\,{\rm Im}\, {\cal L}_{\rm eff}$ at one-loop order in Schwinger's approach.

The second method for defining the decay rate, to be called the Direct or Differential (D) Method is suggested by the fact that the
integrand $\G = 2\,{\rm Im}\, {\cal L}_{\rm eff}$ should be independent of both space and time for persistent external fields of
high symmetry. Then instead of integrating over a large time $T$ one can extract the spacetime volume by identifying the increment
of Fourier modes between $\bf k$ and ${\bf k} + \D \bf k$ that undergo their creation events in each small increment of time
between $t$ and $t + \D t$, in each slice of four-volume between ${\cal V}_4$ and ${\cal V}_4 + \D {\cal V}_4$. The determination of
which Fourier mode(s) undergo a particle creation `event' at each time $t$ effectively establishes a functional relation
${\bf k} = \overline{\bf k}(t)$. Then division by the corresponding four-volume increment gives
\be
\G = \frac{N}{2}\lim_{\D {\cal V}_4 \rightarrow 0}\frac{1}{\D {\cal V}_4}\sum_{\bf k}^{{\bf k} + \D {\bf k}}\, \ln \big(1 + |B_{\bf k}|^2\big)
= \frac{N}{2}\frac{V}{(2 \pi)^3}\int \frac{d^3 {\bf k}}{d{\cal V}_4}\, \ln \big(1 + |B_{\bf k}|^2\big)\Big\vert_{{\bf k} = \overline{\bf k}(t)}
\label{DecayRateDiff}
\ee
for the decay rate in the presence of the persistent background field, due the increment of Fourier modes going through
their pair creation events in an incremental slice of four-volume $d{\cal V}_4$ at $t$, in the limit that both these increments are
infinitesimal. The integration in (\ref{DecayRateDiff}) is to be performed restricted to only those Fourier modes experiencing
a particle creation event at the time $t$, while the values of $|B_{\bf k}|^2$ to be used are still determined by the asymptotic
scattering problem (\ref{asymBog}) for the persistent external field. The differentially defined rate $\G$ will turn out to be
independent of $t$ for persistent fields of high symmetry, thus making the Differential definition (\ref{DecayRateDiff}) equivalent
to the Integral one (\ref{DecayRateInt}) when both methods are applicable, since the time average of a constant is the constant itself.
Since $d {\cal V}_4$ is proportional to $V$, the three-volume factor again will trivially cancel in (\ref{DecayRateDiff}), while this
general form allows a fully covariant definition of the rate in curved space.

The Differential Method of defining the vacuum decay rate (\ref{DecayRateDiff}) has the advantage of avoiding the need
for any background field adiabatic turning on/off and any regulating of the Fourier integral and infinite time limit in (\ref{DecayRateDiff}),
so that there are no non-universal or non-adiabatic edge effects to be concerned with. However, this method requires a more detailed
and explicit characterization of the particle creation events in real time in order to determine $\overline{\bf k}(t)$ and the Jacobian factor
$\frac{d^3 {\bf k}}{d{\cal V}_4}$, not required in the Integral Method, which simply sums all the particle creation taking
place within the full time interval $T$, irrespective of their details at intermediate times.

The characterization of the particle creation `event' needed in the Differential Method is the semi-classical event time, 
$t_{\rm event}({\bf k})$ obtained by inverting ${\bf k} = \overline{\bf k}(t)$, and based upon the behavior of the adiabatic 
phase integral (\ref{genTheta}) in the complex time domain, in particular by the pattern of Stokes and anti-Stokes lines
 of constant Real and Imaginary parts of $\Theta_{\bf k}$ emanating from the complex critical points in $t$ at which 
 $\w_{\bf k}^2$ vanishes. The particle creation `event' is then associated with the time $t$ at which the Stokes' line 
 crosses the real time axis,  as evidenced by the fact that the amplitude of the antiparticle complex conjugate mode function 
 $f_{\bf k}^{(0)*}$ rises rapidly around this time \cite{AndMot1}. Determining the finite subset of Fourier modes that
experience a particle creation `event'  in the finite time interval in this way determines a finite range in the Fourier integral 
in (\ref{DecayRateInt}) proportional to $dt$, which allows the finite rate $\G$ to be determined by (\ref{DecayRateDiff}).

These general considerations and both methods of defining the vacuum decay rate are best illustrated with specific examples.
In this paper we apply both methods to the cases of particle creation in a constant, uniform electric field and in
the gravitational de Sitter background. After first reviewing the persistent field calculation of the vacuum decay and
particle creation, and the complex adiabatic phase method for analyzing particle creation in real time, we present numerical
results on the adiabatic switching on and off again of each background after a long time $T$, and comparison of the vacuum
decay rate $\G$ computed by each method in both cases.

\section{Persistent Uniform Electric Field Background}
\label{Sec:PersistE}

The constant, uniform electric field has been studied by numerous authors by a variety of methods 
\cite{Schw,Nar,Nik,NarNik,FradGitShv,MotPRD85,AndMot1,AndMot2}, and may be considered 
the prototype of the class of problems involving particle creation and quantum vacuum decay of 
classically persistent fields. Choosing the time dependent gauge
\vspace{-4mm}
\be
A_z = -Et\,,\qquad A_t=A_x=A_y = 0
\label{Egauge}
\ee
the Klein-Gordon eq. for a charged scalar field may be separated in Fourier modes as in (\ref{Phiop}) with
\vspace{-4mm}
\be
\w_{\bf k}^2 = (k_z + eEt)^2 + k_{\perp}^2 + m^2  \;.
\label{omegaE}
\ee
We then obtain the mode equation
\vspace{-1mm}
\be
\left[ \frac{d^2}{du^2} + \frac{u^2}{4\,} + \l \right] f_{\l}(u) = 0
\label{yEmode}
\ee
in the dimensionless time and transverse momentum variables
\be
u \equiv \sqrt{\frac{2}{|eE|}} \ \Big(k_z + eEt\Big)\,,\qquad \l \equiv \frac{k_{\perp}^2 + m^2}{2 \, |eE|\,} >  0
\label{ulamdef}
\ee
with $f_{\bf k}(t)$ relabelled as $f_{\l}(u)$. One immediately observes that the frequency function (\ref{omegaE})
is strictly positive everywhere on the real time axis, for $m^2 > 0$.

The function $\d_{\bf k}$ entering the adiabatic condition (\ref{adbcond}) in this case is
\be
\d_{\l} (u) = \frac{1}{2} \frac{(3u^2 - 8\l )}{\,(u^2 +4 \l)^3}
\label{delE}
\ee
from which some properties of particle creation in an electric field background can already be deduced.
First one notices that
\vspace{-2mm}
\be
\d_{\l} (u) \rightarrow  \sdfrac{3\,}{2u^4} \rightarrow 0 \qquad {\rm as} \qquad u \rightarrow \mp\infty
\label{delElims}
\ee
for {\it any} $\l$ or $\bf k$. Thus the adiabatic condition (\ref{adbcond}) is asymptotically satisfied for the persistent,
strictly constant and uniform electric field, in both the $t \rightarrow \mp \infty$ limits, and asymptotic $|0,in\rag$
and $|0,out\rag$ vacuum states exist in which the solutions of (\ref{yEmode}) approach (\ref{adbzero})
and its complex conjugate arbitrarily accurately. This implies in turn that the scattering problem (\ref{asymBog})
is well-posed, and the Bogoliubov coefficients $B_{\bf k}$ are finite and well-defined for each $\bf k$.

In fact the exact solutions of (\ref{yEmode}) are parabolic cylinder functions, whose asymptotic behaviors
and analytic properties are well known. From these solutions and properties one finds \cite{Nar,Nik,NarNik,AndMot1}
\be
B_{\bf k}= -i e^{-\pi\l} = -i\,\exp \left\{- \sdfrac{\pi}{2}\, \sdfrac{k_{\perp}^2 + m^2}{|eE|\,}\right\}
\label{BkE}
\ee
{\it exactly}, for any $\l > 0$. As anticipated by our general discussion in the last section, if this value is
substituted into (\ref{DecayRateInt}) we obtain an indeterminate result, since $|B_{\bf k}|^2$ is independent 
of $k_z$ and the integral over $k_z$ in (\ref{DecayRateInt}) is linearly divergent. This is clearly associated
with the fact that the parallel component of the conserved (canonical) momentum $k_z$ enters the
mode eq. (\ref{yEmode}) together with the time $t$ only through the gauge invariant combination
$k_z + eEt \equiv p_z(t)$, which is the physical kinetic momentum, so that a linear divergence in
$k_z$ is associated with the infinite time in which the $E$-field is applied in obtaining (\ref{BkE}).

Now from (\ref{delE}) the maximum violation of the adiabaticity condition (\ref{adbcond}) is
\be
{\rm max}\, \big\{ |\d_{\l}(u)| \big\}= \frac{1}{16 \l^2} \qquad {\rm at}\qquad u =0, \quad p_z(t)= k_z + eEt = 0
\label{maxdelE}
\ee
so that the time at which this maximum violation occurs is the time
\be
t_{\rm event}(k_z) = -\frac{k_z}{eE} \ \Rightarrow \ \overline k_z(t) = -eEt
\label{tevent}
\ee
when a mode of a given $k_z$ has zero kinetic momentum along the field. By symmetry of (\ref{yEmode}) under
$u \rightarrow -u$ we may expect that (\ref{tevent}) is the time which may be identified with a creation `event' in the mode with
longitudinal canonical momentum $k_z$, and $\overline k_z(t)$ denotes the value of $k_z$ of the Fourier mode
experiencing its creation event at time $t$. Eq. (\ref{tevent}) is the relation between $k_z$ and the time of particle creation
that allows the Jacobian factor
\be
\left\vert \sdfrac{d^3 {\bf k}}{d{\cal V}_4}\right\vert_{{\bf k} = \overline{\bf k}(t)} \!\!\!\!=
\ \sdfrac{d^2 {\bf k}_{\perp}}{V} \left\vert \sdfrac{d\overline k_z}{dt}\right\vert
= \sdfrac{ |eE|}{V}\, d^2 {\bf k}_{\perp}
\label{jacobE}
\ee
appearing in (\ref{DecayRateDiff}) to be computed, eliminating any integral over $k_z$, while the value of ${\bf k}_{\perp}$
is unrestricted and must still be integrated over to give a well-defined result for the vacuum decay rate by the Differential
Method (\ref{DecayRateDiff}). The absolute value must be taken for the Jacobian in order for a positive increment
in $dk_z$ to correspond to a positive increment in time $dt$ for $eE > 0$.

Let us remark also that the maximal violation of adiabaticity in (\ref{maxdelE}) goes to zero as $\l \rightarrow \infty$.
so that, as expected, heavier particles with larger transverse momenta are more difficult to create, however
falling only as a power $\l^{-2}$ for large $\l$, whereas the actual asymptotic value of the created particles in the
mode specified by $\l$ falls exponentially with $\l$, {\it cf.} (\ref{BkE}). The difference in the $\l^{-2}$ power of the
maximum of $|\d_{\l}(u)|$ {\it vs.} the exponential $\l$ dependence of $B_{\bf k}$ illustrates the distinction between
local or transient violations of (\ref{adbcond}) {\it vs.} global or secular particle creation effects which persist at late times.
The asymptotic value of the Bogoliubov coefficients in (\ref{asymBog}) can be obtained by consideration of the
global analyticity properties of the solutions of (\ref{yEmode}), or in the WKB approximation by the behavior of adiabatic
phase (\ref{genTheta}) in the complex time domain \cite{PokKhal,Aud,Fro}.

\begin{figure}[htp]
\vspace{-3mm}
\includegraphics[height=6cm,viewport= 60 0 720 540, clip]{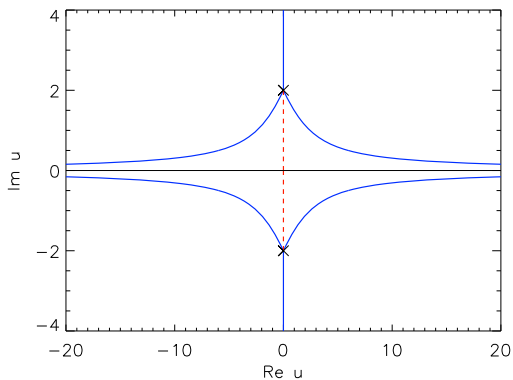}
\vspace{-1.1cm}
\caption{The solid (blue) lines are the three anti-Stokes lines of constant Im $\Theta_\l$ emerging from the two complex zeroes
(\ref{lamzeroes}) of \eqref{omegaE}. The dashed (red) line is the Stokes line of constant Re $\Theta_\l$ connecting
the two critical points, which crosses the real axis at $u=0$, defining the time (\ref{tevent}) of the particle creation event.}
\vspace{-4mm}
\label{Fig:Stokes_Efield}
\end{figure}

The adiabatic phase (\ref{genTheta}) expressed in dimensionless $u, \l$ variables is
\be
\Theta_{\l}(u) = \frac{1}{2}\int^u_0 du\,\sqrt{ u^2 + 4\l}
= \frac{u}{4} \sqrt{u^2 + 4 \l} \,+\, \l \ln \left( \frac{u +  \sqrt{u^2 + 4 \l}}{2 \sqrt\l}\right)
\label{Thetalam}
\ee
in this case, when measured from the symmetric point at $u=0$. Since $\w_k^2 = eE (u- u_{\l}) (u + u_{\l})/2$
has two isolated zeroes in the complex domain, at
\vspace{-1mm}
\be
u= \pm u_{\l} = \pm 2i \sqrt{\l}
\label{lamzeroes}
\ee
where $\w_{\bf k}^2$ vanishes linearly, linear turning point WKB methods may be applied in the
complex domain. From each linear turning point three Stokes lines (of constant Real $\Th_\l$) and three anti-Stokes lines
(of constant Imaginary $\Th_{\l}$) emerge, at $60^\circ$ to each other. The solution of the mode eq. (\ref{yEmode}) that
has the asymptotic limits (\ref{asymBog}) may be found by analytic continuation in the upper half complex $u$ plane
along the solid anti-Stokes lines of constant Im $\Theta_{\l}(u)$ illustrated in Fig. \ref{Fig:Stokes_Efield}.

The square root and logarithm in (\ref{Thetalam}) are defined as the analytic continuation from the real axis of their principal value 
everywhere in the complex $u$ plane, cut by the branch cut taken to be along the positive and negative imaginary axes for $|u| > 2 \sqrt{\l}$.
The constant Im $\Theta_\l$ of the phase function along its anti-Stokes lines in the upper half plane is given by its value at the critical point $+ u_{\l}$, 
\vspace{-5mm}
\be
{\rm Im}\, \Theta_{\l}(u_{\l}) = {\rm Im}\, \big[\l \ln (i)\big] = \frac{\pi\l}{2}
\label{ImThetalam}
\ee
and the adiabatic mode function (\ref{adbzero}) is a good approximation to the exact solution everywhere along the $u$ contour defined
by the solid blue anti-Stokes line in the upper half-plane, except in the vicinity of the complex turning point $u= u_{\l}$. There a 
standard WKB linear turning point analysis and matching of the asymptotic solutions on the two halves of the anti-Stokes contour determines
\be
B_{\l} = -i \exp \big[- 2\, {\rm Im} \, \Theta_{\l}(u_{\l})\big] = -i\,e^{-\pi\l}
\label{Btotlam}
\ee
which coincides with the exact value (\ref{BkE}) in this simple example of only one linear complex critical point in the
upper half $u$-plane.

\begin{figure}[htp]
\vspace{-7mm}
\includegraphics[height=6cm,viewport= 0 0 760 490, clip]{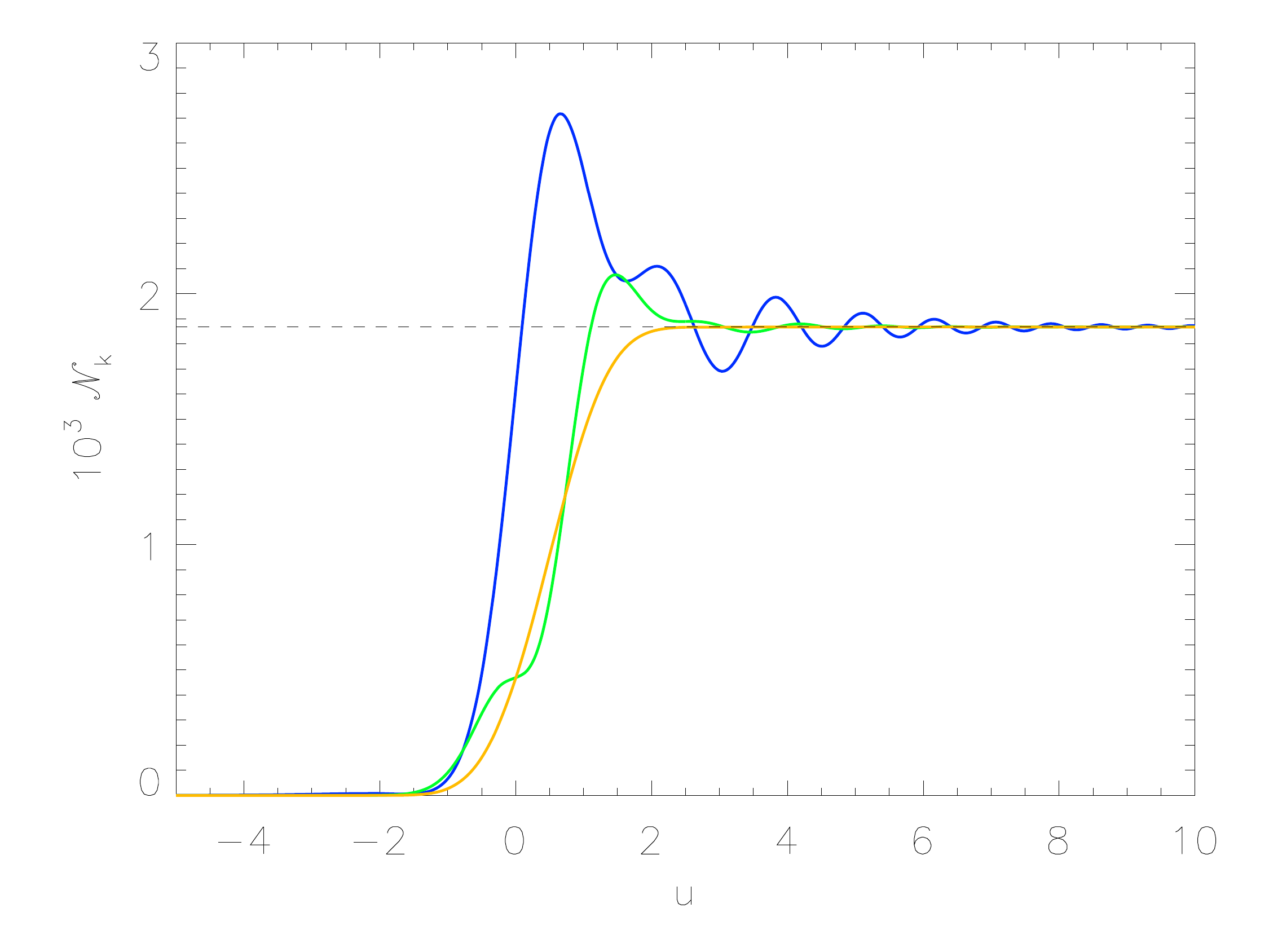}
\vspace{-6mm}
\caption{The mean number of particles created from the vacuum by a constant, uniform electric field as a function of rescaled time $u$, for $\l =1$. 
The first two curves (blue and green) are the adiabatic particle numbers ${\cal N}_{\bf k}^{(n)}$, defined by (\ref{adbpart}) for $n=1,2$ with $f_{\bf k}= f_{\l\,(+)}(u)$ 
the $in$-state solution of (\ref{yEmode}), given by Eqs.~(4.9a) and (5.11)-(5.13) of Ref.~\cite{AndMot1}. The third curve (orange) with no intermediate maxima 
and minima is the `superadiabatic' particle number (\ref{superadbE}) \cite{Ber,DabDun}. Aside from transient effects dependent on the adiabatic order, all three curves rise 
rapidly from zero in the vicinity of $u=0$ and tend to the same asymptotic value $e^{-2 \pi}= 1.86744 \times 10^{-3}$ of (\ref{Blamsq}) as $u\rightarrow \infty$.}
\vspace{-4mm}
\label{Fig:EW0W2}
\end{figure}

The number density of particles in momentum $\bf k$ as $t\rightarrow \infty$, if started in the initial state vacuum
at $t \rightarrow -\infty$ is therefore
\be
|B_{\l}|^2 = \exp\big[- 4\, {\rm Im}\, \Theta_{\l}(u_{\l})\big] =
e^{-2\pi\l} = \exp\left[- \frac{\pi (k_{\perp}^2 + m^2)}{eE} \right]
\label{Blamsq}
\ee
and the solutions of the mode eq. (\ref{yEmode}) exhibit a fairly sharp transition illustrated in Fig.~\ref{Fig:EW0W2} from the early to late time asymptotic forms
(\ref{asymBog}) at $u=0$ where the Stokes line of constant Re $\Theta_{\l}$ crosses the real time axis at $u=0$ in Fig.~\ref{Fig:Stokes_Efield}.

The particle creation event is defined by the rapid rise in adiabatic particle number defined by (\ref{adbpart}), together with Eqs.~(4.9a) and 
(5.11)-(5.13) of Ref.~\cite{AndMot1}, and illustrated in Fig.~\ref{Fig:EW0W2}.\footnote{The first line of Eq. (5.13) of Ref.~\cite{AndMot1} 
contains a typographical error in its last term which should read $\frac{3}{8} \frac{\dot\w_{\bf k}^2}{\w_{\bf k}^3}$.} For comparison the universal, 
optimally adiabatic, or {\it `superadiabatic'} particle number $\overline {\cal N}_\l$ \cite{Ber,DabDun}, given in terms of the adiabatic phase integral by
\be
\overline {\cal N}_\l(u) = \frac{1}{4}\exp\!\big[\!- 4\, {\rm Im}\, \Theta_{\l}(u_{\l})\big]\left\{ {\rm erfc}\! \left[\frac{-\Theta_\l(u)}{\sqrt{{\rm Im}\,
\Theta_{\l}(u_{\l})}}\right]\right\}^2 =  \frac{e^{-2 \pi \l}}{4}\left\{ {\rm erfc}\! \left[-\sqrt{\sdfrac{2}{\pi \l}}\, \Theta_\l(u)\right]\right\}^2
\label{superadbE}
\ee
is also shown in Fig.~\ref{Fig:EW0W2} for the $E$-field. The time of the event at $u=0$ coincides in this case with the maximum value of $|\d_{\l}(u)|$, 
(\ref{maxdelE}). Aside from non-universal transients illustrating the quantum uncertainty in defining particle number at the transition, which depend 
upon the definition of particle number used, the particle creation event is characterized by a permanent secular rise (\ref{Blamsq}). This asymptotic 
particle number is unambiguously defined and independent of adiabatic order, but exponentially small in $\l$ for large $\l$, and can be obtained 
from the global analysis of the adiabatic phase (\ref{Thetalam}), and its critical point $u_\l$ (\ref{lamzeroes}) in the complex time domain.

This detailed understanding of the Stokes' lines of the adiabatic phase and time (\ref{tevent}) when each $k_z$ mode goes through
its creation event determines the Jacobian factor (\ref{jacobE}) in the differential rate formula (\ref{DecayRateDiff}) for the constant
$E$-field background. Equivalently it also informs us how to regulate the $k_z$ integral in a finite time $T$ in the integral
formula (\ref{DecayRateInt}). For if one starts in the adiabatic vacuum with mode function (\ref{adbzero}) for all modes
at some {\it finite} initial time $t_0$, only those modes for which
\vspace{-3mm}
\be
p_z(t_0) <0:\ t_{\rm event}(k_z)  > t_0\qquad {\rm but} \qquad p_z(t_1) > 0: \ t_{\rm event}(k_z) < t_1
\ee
experience their particle creation event between $t_0$ and $t_1$. Thus we may approximate
\vspace{-2mm}
\be
\vert B_{\bf k}\vert^2  \simeq   \left\{ \begin{array}{l} e^{-2\pi \l}\quad {\rm for} \quad -eEt_1 < k_z < -eEt_0\\
0 \qquad\  \ \ {\rm otherwise} \end{array} \right.
\label{Bwindow}
\ee
and in the finite elapsed time $T= t_1-t_0$ only modes in the $k_z$ interval of the window linearly growing in time in (\ref{Bwindow})
give a non-vanishing contribution to the vacuum decay rate. With the step function approximation of (\ref{Bwindow}), (\ref{inout}) and
(\ref{DecayRateInt}) then yield
\begin{align}
\G &= \lim_{T\rightarrow \infty}\frac{1}{T} \int_{-eEt_1}^{-eEt_0}\frac{dk_z}{2 \pi} \int \frac{d^2{\bf k}_{\perp}}{(2 \pi)^2} \,\ln\, \big(1 +  e^{- 2 \pi \l}\big)\nn
&=  \frac{eE}{2\pi} \int \frac{d^2{\bf k}_{\perp}}{(2 \pi)^2}\,   \ln\left\{1 + \exp\left[- \frac{\pi\, (k_{\perp}^2 + m^2)}{eE}\right]\right\}\nn
&=  \frac{eE}{2(2\pi)^2} \int_0^{\infty} dk^2_{\perp}\, \sum_{n=1}^{\infty} \frac{(-)^{n+1}\hspace{-6mm}}{n}
\quad \exp\left[-\frac{\pi n\, (k_{\perp}^2 + m^2)}{eE}\right]\nn
&=\frac{(eE)^2}{(2\pi)^3}\, \sum_{n=1}^{\infty} \frac{(-)^{n+1}\hspace{-6mm}}{n^2} \
\hspace{4mm}\exp\left(-\frac{\pi n  m^2}{eE}\right)
\label{vacdecayE}
\end{align}
which agrees with Schwinger's proper time method for the calculation of the decay rate of a uniform electric field into scalar
particle/antiparticle pairs \cite{Schw}. Clearly the identical expression is obtained from the differential formula (\ref{DecayRateDiff})
upon making use of (\ref{jacobE}) which eliminates the $k_z$ integral, $T$ dependence and limit entirely, giving directly the second
line of (\ref{vacdecayE}). We next verify (\ref{vacdecayE}) by turning the $E$-field on and off adiabatically, letting it last for
a very long time $T$ and extrapolating to the limit indicated in the Integral Method (\ref{DecayRateInt}) numerically.

\section{Adiabatic Switching On/Off of a Uniform Electric Field}
\label{Sec:Eadb}

Before discussing the $E$-field profile needed to compute the vacuum decay rate by the Integral Method, we mention first
the modified $E$-field time profile
\be
{\bf E}(t) = E\, {\bf \hat z}\, \sech^2\, (t/T)
\label{Esech}
\ee
that has been considered in the literature \cite{Sau,GavGit}, for which the electric field vanishes asymptotically in both the $t \rightarrow \mp \infty$ limits.
This corresponds to the spatially uniform gauge potential
\be
A_z(t) = -ET \tanh \,(t/T)
\label{Aztanh}
\ee
for which the mode eq. (\ref{harmosc}) with
\be
\w_{\bf k}^2 (t) \equiv \big[k_z + eET \tanh\,(t/T)\big]^2 + k_{\perp}^2 + m^2
\label{freqT}
\ee
may be solved exactly in terms of hypergeometric functions \cite{Sau,GavGit}.  The frequency has the asymptotic limits
\vspace{-2mm}
\be
\lim_{t\rightarrow \pm \infty} \w_{\bf k} (t) \equiv \w_{\bf k}^{\pm} = \sqrt{(k_z \pm  eET)^2 + k_{\perp}^2 + m^2}
\ee
which are constants. Thus the positive frequency particle and negative frequency antiparticle modes are the unique zero-field modes in
each asymptotic limit. From the analytic properties of the exact hypergeometric function solutions of the mode equation (\ref{harmosc})
with (\ref{freqT}), the Bogoliubov coefficients of the scattering problem (\ref{asymBog}) may also be computed analytically, with the
result \cite{Sau,Nik,GavGit}
\be
|A_{\bf k}(T)|^2 = \frac{\bigg.\cosh^2 \left[ \dfrac{\pi}{2}\sqrt{(2eET^2)^2 - 1}\right] +
\sinh^2\left[\dfrac{\pi}{2} \, (\omega_{\bf k}^+ + \omega_{\bf k}^-)\,T\right]\bigg.}
{\Big. \sinh (\pi \omega_{\bf k}^+ T)\ \sinh (\pi\omega_{\bf k}^-T)\Big.}
\ee
\be
|B_{\bf k}(T)|^2 = \frac{\bigg.\cosh^2 \left[ \dfrac{\pi}{2}\sqrt{(2eET^2)^2 - 1}\right] +
\sinh^2\left[\dfrac{\pi}{2} \, (\omega_{\bf k}^+ - \omega_{\bf k}^-)\,T\right]\bigg.}
{\Big. \sinh (\pi \omega_{\bf k}^+ T)\ \sinh (\pi\omega_{\bf k}^-T)\Big.}
\label{BkadbT}
\ee
satisfying (\ref{Bogcond}) for all $\bf k$ and $T$.

\begin{figure}[htp]
\vspace{-1.5cm}
\includegraphics[height=6cm,viewport= 0 10 760 600, clip]{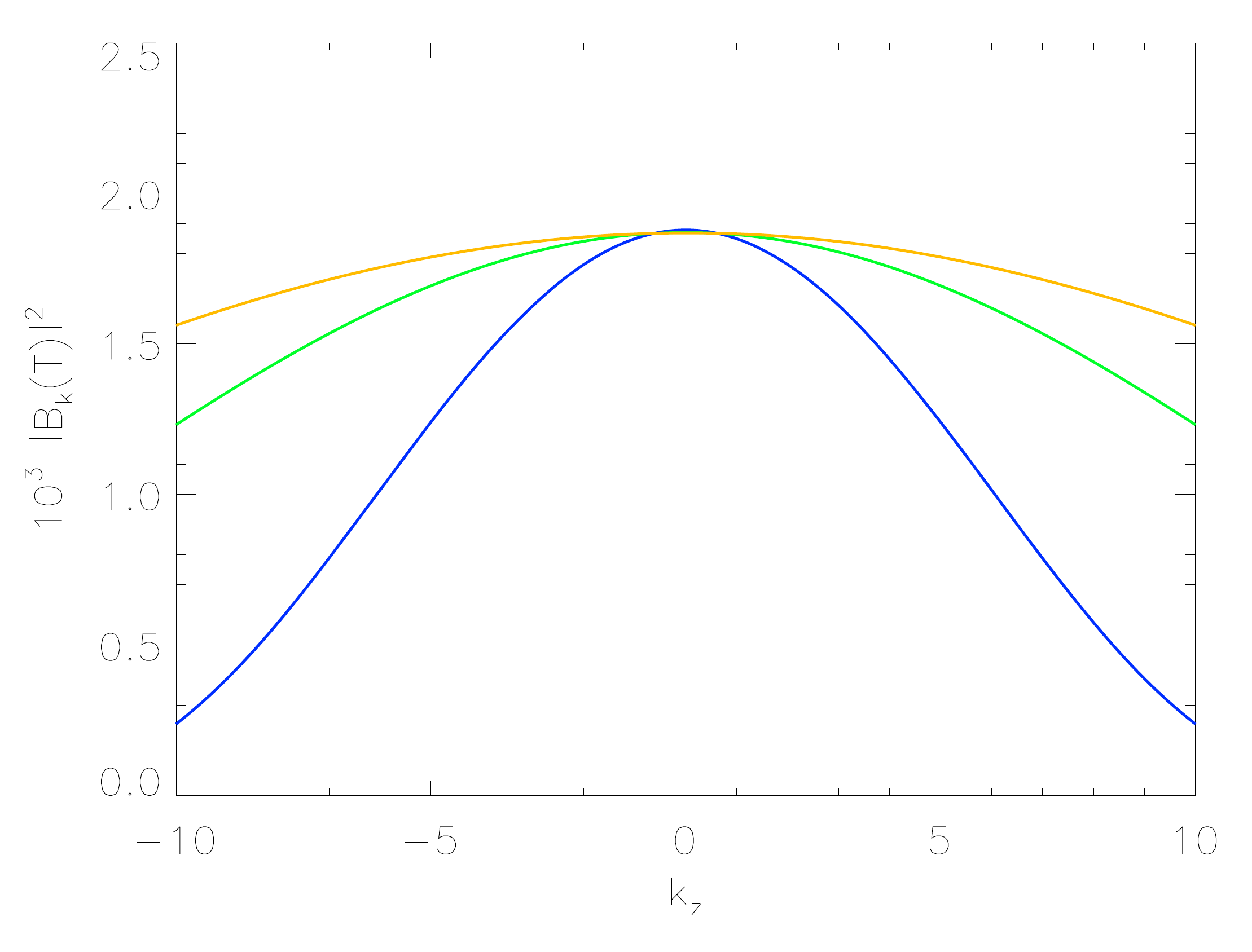}
\vspace{-5mm}
\caption{Particle creation $|B_{\bf k}(T)|^2$ given by (\ref{BkadbT}) for the $E$-field (\ref{Esech}), as a function of $k_z$ for $\l = 1$. 
The curve that falls off the most rapidly in $|k_z|$ (blue) is for $\sqrt{eE} \, T= 20$, the middle one (green) is for $\sqrt{eE}\, T = 40$, 
and the outer one (orange) is for $\sqrt{eE}\, T = 60$, showing the approach to $e^{-2 \pi \l}$ near $k_z=0$.}
\vspace{-4mm}
\label{Fig:B21paramEfield}
\end{figure}

The Bogoliubov coefficient $|B_{\bf k}(T)|^2$ is well-behaved as $|k_z| \rightarrow \infty\ (T$ fixed), being proportional
to $\exp(-2 \pi |k_z| T)$ and vanishing exponentially in that limit. Hence for any finite $T$ the integral over $k_z$ and
the total number of particles created is finite. In the opposite limit, with $k_z$ fixed
\be
\lim_{T \rightarrow \infty} \, |B_{\bf k}(T)|^2 = \exp \left(-\pi\, \sdfrac{k_\perp^2 + m^2}{eE}\right) = e^{-2 \pi \l}
\label{Bsqlim}
\ee
the value in the constant electric field (\ref{Blamsq}) independent of $k_z$ is recovered. Thus the large $k_z$ and large $T$ limits
of (\ref{BkadbT}) do not commute. The behavior of $|B_{\bf k}(T)|^2$ as a function of $k_z$ and of $T$ is shown in
Figs. \ref{Fig:B21paramEfield}, with the flattening for small $|k_z|$ as $T \rightarrow \infty$ according to (\ref{Bsqlim}) illustrated.

The time profile (\ref{Esech}) cannot be used to compute the decay rate of a constant $E$-field by the Integral Method
(\ref{DecayRateInt}), because the $E$-field (\ref{Esech}) is not constant over times of order $T$. Although the turning on and
off of the $E$-field in (\ref{Esech}) is adiabatic, the time for the transition also grows with $T$ and hence
as Fig. \ref{Fig:B21paramEfield} shows, the particle production (\ref{BkadbT}) falls off smoothly in $|k_z|$ rather than sharply
outside a well-defined window in $k_z$, as required to match the behavior (\ref{Bwindow}) for a constant $E$-field.
However we may extract the constant $E$-field Schwinger rate from (\ref{Esech})-(\ref{Bsqlim}) through
the Differential Rate formula (\ref{DecayRateDiff}), provided we compute the Jacobian (\ref{jacobE}), restricted to finite
values of $k_z$ in the limit $T \rightarrow \infty$, where (\ref{Bsqlim}) holds, corresponding to finite times $|t| \ll T$ when
the $E$-field (\ref{Esech}) is constant. In that limit because of (\ref{Bsqlim}) and making use of the Jacobian (\ref{jacobE}) based on
the constant $E$ limit, the second line of (\ref{vacdecayE}) and finally Schwinger's vacuum decay rate for a constant $E$-field
is recovered.

\begin{figure}[htp]
\begin{center}
\begin{tabular}{ll}
\hspace{-1cm}
\includegraphics[scale=0.5,height=6cm,viewport= 60 0 750 540, clip]{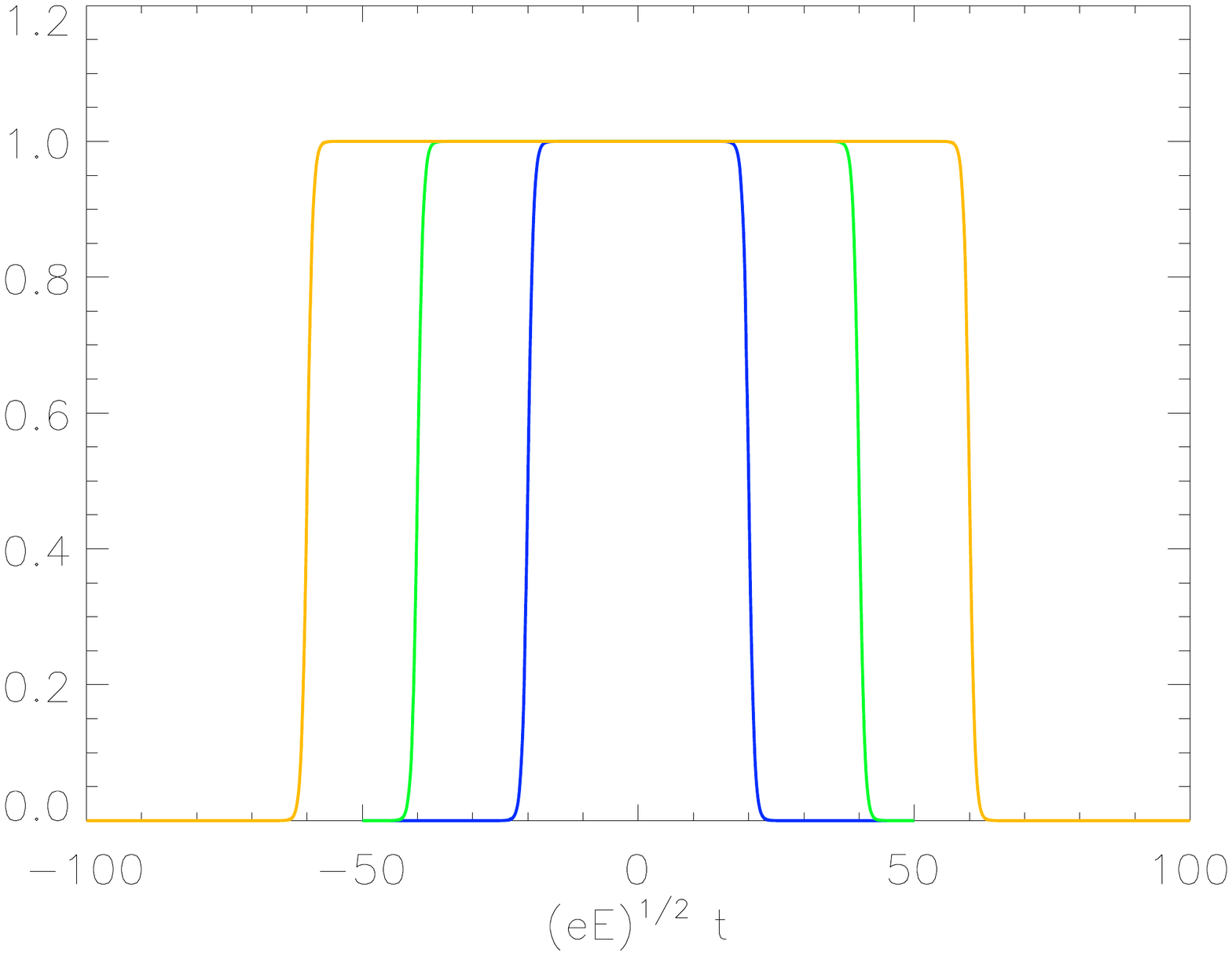}
\hspace{2mm}
\includegraphics[scale=0.5,height=6cm,viewport= 60 0 750 540, clip]{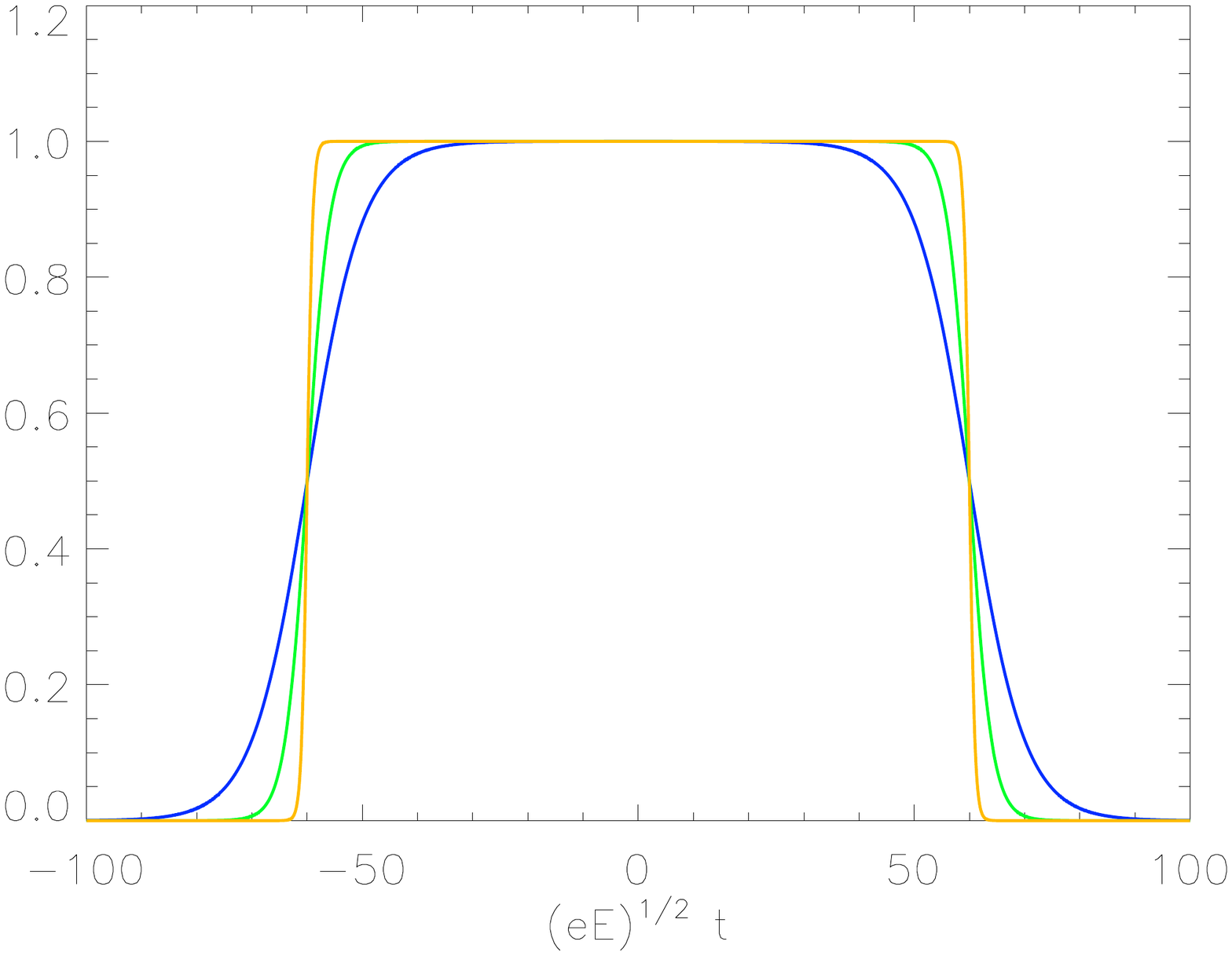}
\vspace{-5mm}
\end{tabular}
\end{center}
\vspace{-1cm}
\caption{The electric field for the profile \eqref{Eprofile} in units of its maximum as a function of time with $t_1 = -t_0$.
The curves in the left panel show fixed $b/\sqrt{e E} =1$, with $\sqrt{eE}\, t_1 = 20$ (blue), $ 40$ (green),  $60$ (orange). The curves in the right
panel show fixed $\sqrt{eE}\, t_1  = 60$ with $b/\sqrt{eE} = 0.1$ (blue), $ 0.25$ (green), $1$ (orange).}
\vspace{-5mm}
\label{Fig:Eprofile}
\end{figure}

In order to use the Integral Method (\ref{DecayRateInt}) one needs instead at least a two parameter family of time profiles in which
the parameter controlling the duration of the field is separate and distinct from the parameter controlling the time during
which the field is switched on and off again. An analytical function with these properties is
\be
{\bf E}(t) = \frac{E\,{\bf \hat z}}{2} \Big\{ \tanh \big[b\,(t-t_0)\big] -  \tanh \big[b\,(t-t_1)\big]\Big\}
\label{Eprofile}
\ee
some examples of which are shown in Fig. \ref{Fig:Eprofile}. This profile has the property that $E(t)$ vanishes well before some initial time $t_0$
and well after some final time $t_1$ where $t_1-t_0 \equiv T > 0$. Now $T$ can be taken arbitrarily large, while ${\bf E}(t)$ has approximately the constant
value $E\, {\bf \hat z}$ between $t_0$ and $t_1$, and is adiabatically switched on and off on an arbitrary time scale of order $b^{-1}$: {\it cf.} Fig. \ref{Fig:Eprofile}.
Thus if $b$ is small enough, the particle creation during the adiabatic switching on and off of the $E$-field may be kept small, and rendered
negligible compare to the particle creation during the arbitrarily long interval of time $T$ when the field is constant.

\begin{figure}[htp]
\vspace{-9mm}
\includegraphics[height=6cm,viewport= 50 50 760 600, clip]{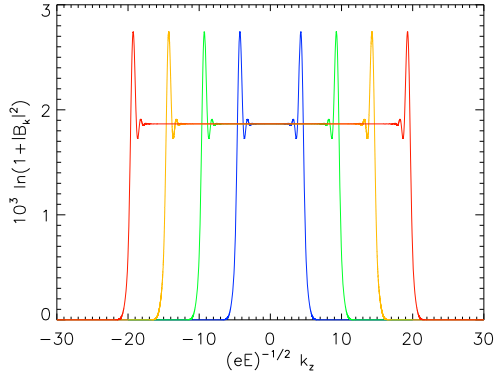}
\vspace{-6mm}
\caption{The numerically computed integrand $\ln (1+|B_{\bf k}|^2)$ for the rate as a function of $k_z$  for four different values of $t_1$
when $\l= b/\sqrt{eE} = 1$ and $t_0 = -t_1$. Going out from $k_z = 0$ in either direction the curves correspond to $\sqrt{eE} t_1  = 10$ (blue), 20 (green)
30 (orange), and 40 (red).}
\vspace{-1mm}
\label{Fig:Ebetas}
\end{figure}

The gauge potential corresponding to (\ref{Eprofile}) may be taken to be
\be
A_z(t) = -\frac{E}{2b} \, \ln\left\{ \frac{\cosh \left[b\,(t-t_0)\right]}{\cosh \left[b\,(t-t_1)\right]}\right\} - \frac{E(t_0+t_1)}{2}
\label{Aprofile}
\ee
which behaves as
\vspace{-5mm}
\be
A_z(t) \simeq -E\,  \left\{\begin{array}{cc} \,t_0 & \quad t \ll t_0 \\
t & \quad t_0 \ll t \ll  t_1 \\
\,t_1 &\quad  t_1 \ll t
\end{array}\right.
\label{Aztlim}
\ee
for $b(t_1-t_0)= bT \gg 1$. For this potential no analytic solution for the mode eq. (\ref{harmosc}) is known and we must rely on
a numerical solution. The Bogoliubov coefficients $|B_{\bf k}|^2$ are finite as is the integral over all modes, and the decay rate is
now computed by the Integral Method (\ref{DecayRateInt}), taking the $T \rightarrow\infty$ limit numerically. The numerical results
for the integrand $\ln (1 + |B_{\bf k}|^2)$ shown in  Fig. \ref{Fig:Ebetas} (unlike Fig. \ref{Fig:B21paramEfield})
now show the expected linear opening of the approximately constant window function in $k_z$ as $T= t_1-t_0$ is increased.

\begin{figure}[htp]
\vspace{-1.2cm}
\includegraphics[height=6cm,viewport= 50 50 760 600, clip]{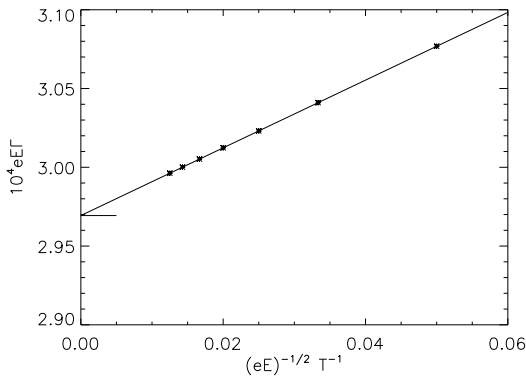}
\vspace{-5mm}
\caption{The two dimensional rate (\ref{Gamma2D}) as a function of $T^{-1}$ for the case $\l = b/\sqrt{eE} = 1$.   The crosses are our numerical data.
The solid line going through them is a least squares fit to the data which is extrapolated to $T^{-1}\rightarrow 0$. The horizontal line segment gives 
the value of the rate for this value of $\l$ when the electric field is static, according to the Schwinger formula (\ref{vacdecayE}), to which the numerical 
results extrapolate.}
\vspace{-4mm}
\label{Fig:ErateT}
\end{figure}

For a uniform $E$-field in two spacetime dimensions, dropping the $d^2{\bf k}_{\perp}/(2\pi)^2$ integral, the Integral Rate \eqref{DecayRateInt} is
\vspace{-4mm}
\be
\G_{2D}= \lim_{T\rightarrow \infty}  \frac{1}{T} \int_{-\infty}^\infty \frac{d k_z}{2 \pi} \ln(1+|B_\mathbf{k}|^2)  \;.
\label{Gamma2D}
\ee
which is shown as a function of $T$ in Fig. \ref{Fig:ErateT}, with the limit extrapolated to the Schwinger result in $d=2$ for $T^{-1} = 0$.
The linear fit to the $1/T$ extrapolation shows that the finite edge effects and particle creation due to the switching on and off of the $E$-field 
around $t=t_0$ and $t=t_1$ remain finite while the constant $E$-field contribution to (\ref{Gamma2D}) increases linearly as the time interval 
$T$ increases.

\begin{figure}[htp]
\vspace{-7mm}
\begin{center}
\begin{tabular}{ll}
\includegraphics[height=6cm,viewport= 100 60 860 535, clip]{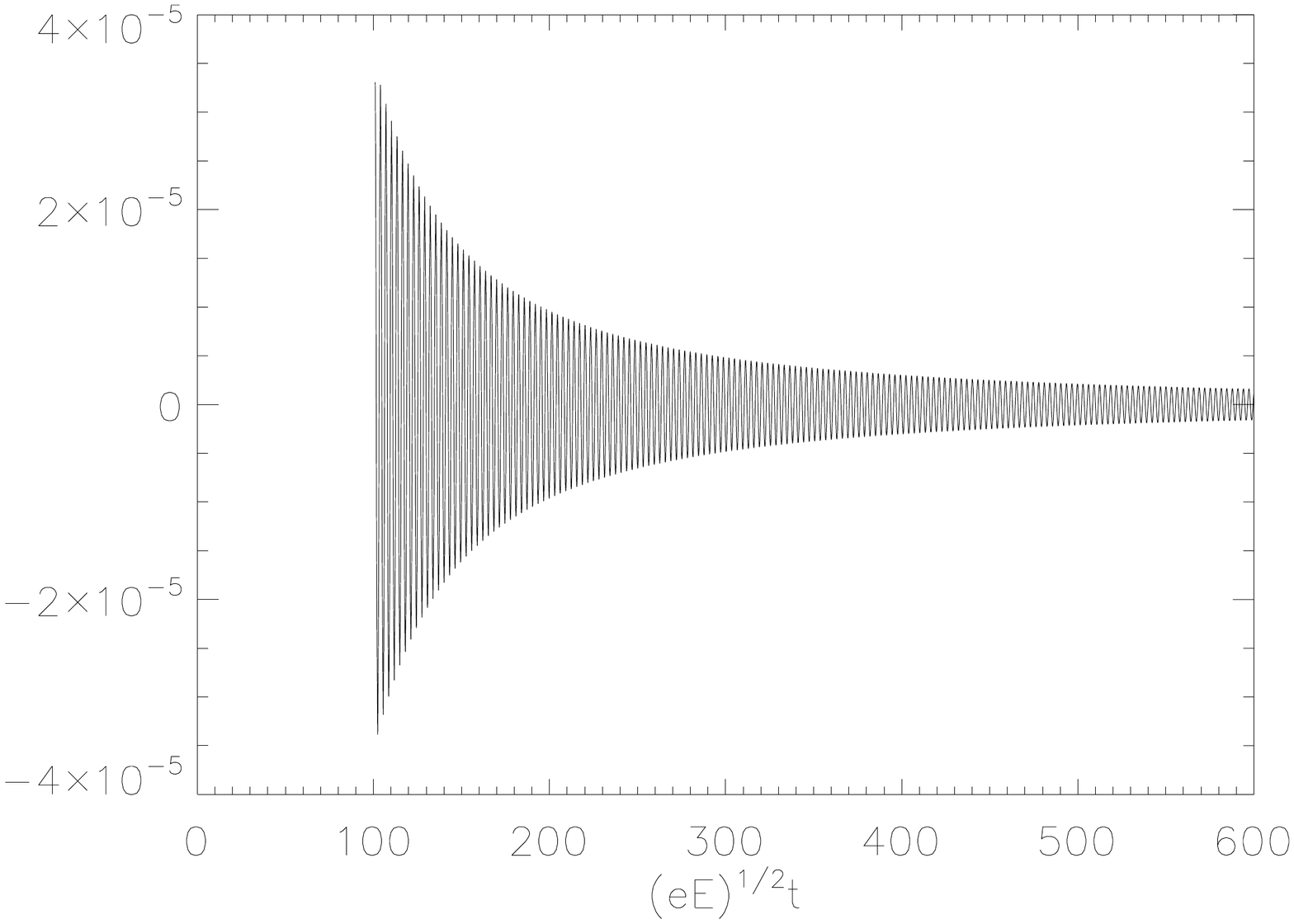}
\hspace{-1.5cm}
\includegraphics[height=6cm,viewport= 100 60 860 535, clip]{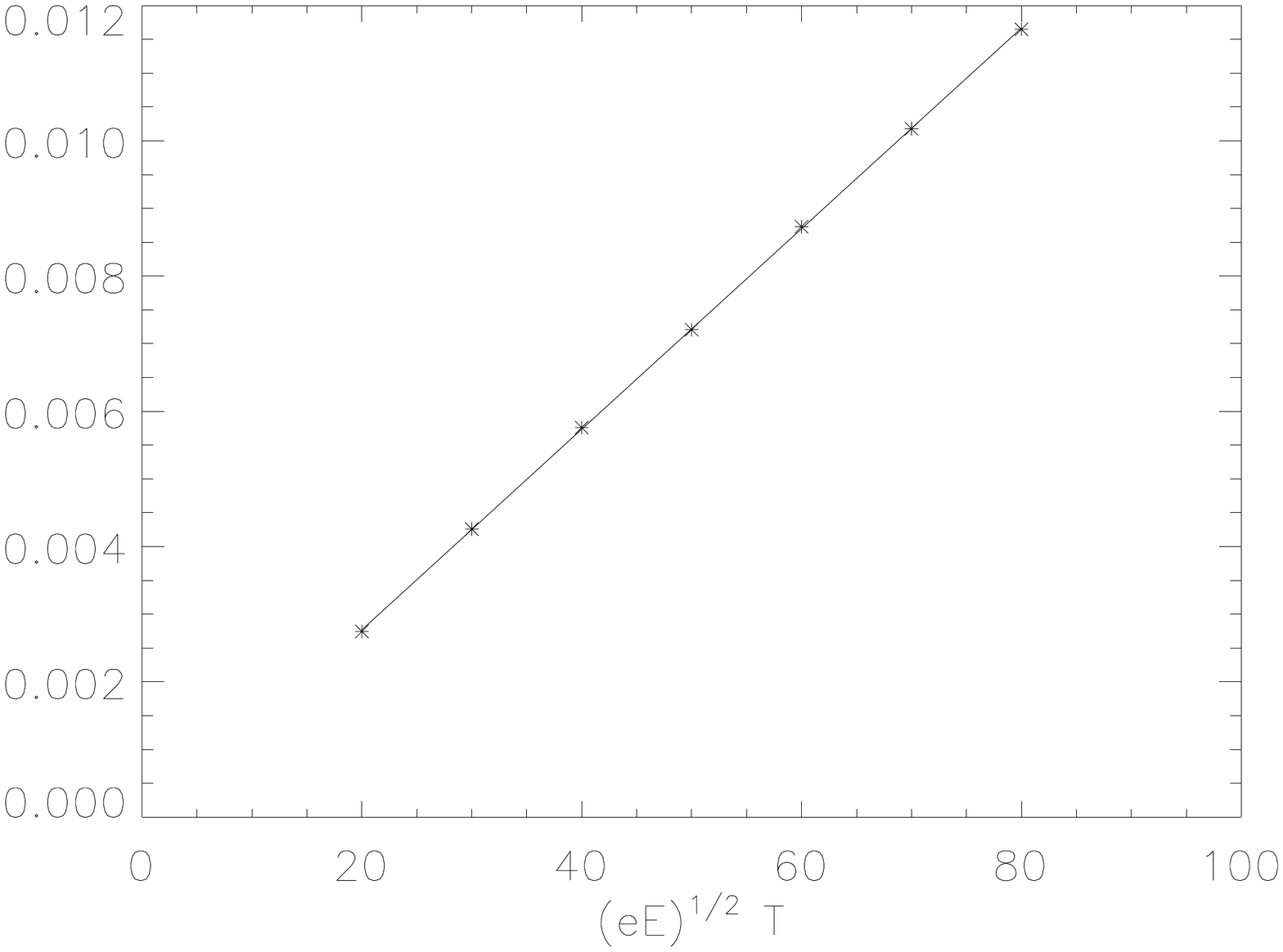}
\vspace{-7mm}
\end{tabular}
\end{center}
\caption{Left Panel: The oscillating $A_{\bf k} B^*_{\bf k}$ part of the current (\ref{jz}) with $\l = 1$ as a function of time for the case $t_1 = -t_0 = 30 (eE)^{-1/2}$
corresponding to $T = 60 (eE)^{-1/2}$, showing that it averages to zero and that its  oscillations are damped at late times after the $E$-field is turned off.
Right Panel: The constant non-oscillating $|B_{\bf k}|^2$ particle contribution to the current (\ref{jz}) with $\l = 1$ at late times,
as a function of the time $T$ for which the $E$-field was turned on. The solid line is a linear $T$ fit to the numerical data. }
\vspace{-4mm}
\label{Fig:jzoscT}
\end{figure}

The electric current from the created charged particle pairs at the end of the process, {\it i.e.} at late times after the electric field has been turned
off, is easily evaluated.  In this case, the vacuum-subtracted $z$ component of the electric current the two-dimensional case at late times is
\bea
&&\lag j_z \rag\Big\vert_{d=2} =  2e\! \int_{-\infty}^\infty \frac{d k_z}{2\pi} \, (k_z - eA_z) \left[ |f_{\bf k}(t)|^2 - \frac{1}{2 \w_{\bf k}}\right]\nn
&&\hspace{-7mm}=  2e\! \int_{-\infty}^\infty \frac{d k_z}{2\pi} \, \frac{(k_z - eA_z)}{\w_{\bf k}}
\, \Big[ |B_{\bf k}|^2 + {\rm Re} \left(A_{\bf k} B^*_{\bf k} e^{-i\w_{\bf k}t} \right) \Big]
\rightarrow 2e\! \int_{-\infty}^\infty \frac{d k_z}{2\pi} \, \frac{(k_z - eA_z)}{\w_{\bf k}}\, |B_{\bf k}|^2
\label{jz}
\eea
which contains a non-oscillating constant $|B_{\bf k}|^2$ term from the created particles, as well as a rapidly oscillating quantum interference $A_{\bf k} B^*_{\bf k}$ 
term. The latter gives rise to a rapidly oscillating transient contribution that decays away with time due to phase cancellations (Left Panel of Fig. \ref{Fig:jzoscT}). 
In contrast the contribution from the created particles gives a constant contribution to the current at late times, whose value depends linearly upon the total time $T$ 
during which the electric field was on, as expected from the linerarly opening window in $k_z$ (Right Panel of Fig. \ref{Fig:jzoscT}). This linear growth with $T$ 
shows that the backreaction of the current of the particles created by a persistent electric field must eventually be taken into account, no matter how small the 
coupling $e$ \cite{KESCM}.

\section{Vacuum Decay of De Sitter Space: Flat Spatial Sections}
\label{Sec:dSPersist}

The globally complete closed ${\mathbb S}^3$ spatial sections and contracting part of de Sitter space have been considered in detail in \cite{AndMot1,AndMot2},
showing that global de Sitter space is unstable. Here we specialize to the flat spatial sections of spatially homogeneous and isotropic 
Friedman-Lemaitre-Robertson-Walker (FLRW) spacetimes with the metric line element
\vspace{-1mm}
\be
ds^2 = -dt^2 + a^2(t)\, d{\bf x}^2
\label{dSflat}
\ee
presumed most relevant for cosmology. In the FLRW background geometry (\ref{dSflat}) the scalar wave eq.~$(-\sq + M^2) \Phi =0$ separates, and the 
scalar field may be represented as a Fourier sum analogous to (\ref{Phiop}), with mode solutions of the form $\Phi \sim \phi_k(t) e^{i {\bf k \cdot x}}$. Removing 
the scale factor by defining the complex mode function $f_k(t) = a(t)^{\frac{3}{2}} \phi_k(t)$ gives a mode equation for $f_k(t)$ which is 
again of the form (\ref{harmosc}) with a time dependent frequency
\be
\w_{\bf k}^2(t) = \sdfrac{{\bf k}^2}{a^2} + m^2  - \sdfrac{h^2}{4\,} - \sdfrac{\dot h}{2}  + (6\xi -1) (2 h^2 + \dot h)
\label{FLRWomega}
\ee
where $h\equiv \dot a/a$ for general $a(t)$.  We consider here the case of conformal coupling $\xi = \frac{1}{6}$ to simplify the algebra,
although the same methods may be applied for any $\xi$.  For de Sitter space $a(t)= a_{\rm dS}(t) = \exp(H t),$ with $h=H$ a constant and $\dot h = 0$.
Then defining the dimensionless time variable $u\equiv H t$ and dimensionless momentum $k \equiv |{\bf k}|/H$, the oscillator equation (\ref{harmosc})
becomes
\be
\left[ \frac{d^2}{du^2} + \w^2_{k\g} (u)\right] f_{k\g}(u) =0
\label{modeqflat}
\ee
with the time dependent dimensionless frequency function given by
\be
\w^2_{k\g} (u) = k^2 e^{-2u} + \g^2  \qquad {\rm and\ with} \qquad  \g \equiv \sqrt{ \sdfrac{m^2}{H^2} - \sdfrac{1}{4}}\ .
\label{omflat}
\ee
We restrict ourselves here to the massive case $m^2 > H^2/4$, in the principal series spin-$0$ representation of the $SO(4,1)$
de Sitter isometry group \cite{BorDur}, so that $\g$ is real and positive, as is $\w^2_{k\g}(u)$.  

The adiabatic parameter appearing in (\ref{adbcond}) in this case is
\be
\d_{k\g}(u) = \sdfrac{k^2 e^{-2u} }{8\,\omega^6_{k\g}}\, \left( k^2 e^{-2u}- 4 \g^2\right)
= \sdfrac{1}{8\,\omega^2_{k\g}}\left( 1- \sdfrac {\g^2\,}{\w_{k\g}^2}\right) \left( 1- \sdfrac {5\,\g^2}{\w_{k\g}^2}\right)
\label{deldS}
\ee
which reveals that as in $E$-field case (\ref{delElims}), so also in de Sitter space
\be
\lim_{u \rightarrow \mp \infty} \d_{k\g}(u)  = 0\qquad  {\rm for\ every}  \quad  k, \g \ge 0\,.
\label{ulims}
\ee
Hence there is a well-defined adiabatic $|0,in\rag$ and $|0, out\rag$ vacuum state asymptotic in each infinite time limit of de Sitter space
the scattering problem (\ref{asymBog}) is again well-posed, and the Bogoliubov coefficients $B_{\bf k}$ finite and well-defined
for every $\bf k$. In between the asymptotic limiting times (\ref{ulims}), at a finite $u$ of order $\ln(k/\g)$ the absolute value of $|\d_{k\g}|$ attains the maximum 
\be
{\rm max} \, |\d_{k\g}|  \simeq \sdfrac{0.0656423}{\g^2}
\ee
which may be compared to (\ref{maxdelE}) in the $E$-field case. Around this time we may expect the given $k$ mode to experience a creation event.

Determining the correct magnitude of the secular particle creation effect and its detail in real time again requires a global analysis of
the adiabatic phase integral in the complex time domain. Changing variables to the physical momentum (in units of $H$)
\be
z \equiv \sdfrac{k}{a} = k \, e^{-u}
\label{zdef}
\ee
so that $\w_{k\g} = \sqrt{z^2 + \g^2}$, one finds the adiabatic phase integral
\bea
\Th_{\g}(z) &\equiv &\int^{u(z)}_{u_{k\g}}du\ \w_{k\g}(u)= - \int^z_{\g\ka} \frac{dz}{z} \sqrt{z^2 + \g^2}\nn
&=& - \sqrt{z^2 + \g^2} + \g \ln \left[ \frac{\sqrt{z^2 + \g^2} + \g}{z}\right]
\rightarrow  \left\{\begin{array}{cr} \ \ \ -z \ \ \quad\quad z\rightarrow \infty\\ -\g\ln z \,\qquad z \rightarrow 0^+ \end{array}\right.
\label{ThetadS}
\eea
for the flat Poincar\'e sections of de Sitter space. The lower limit of integration has been set so that $\Th_\g = 0$ at $z=\ka\g$, with the
corresponding $u_{k \g}$ the time at which the Stokes line crosses the real axis: {\it cf.}  \eqref{ukgamflat} below. This
$\ka$ is defined therefore as the solution of the transcendental equation
\be
\sqrt{\ka^2 + 1} = \ln \left[\frac{\sqrt{\ka^2 + 1} + 1}{\ka}\right] \ \Rightarrow\  \ka\simeq 0.662743 \;.
\label{kapdef}
\ee
As in the constant $E$-field case, the solutions of the mode eq.~({\ref{omflat}) in persistent or `eternal' de Sitter space are known analytically. The
change of variable to $z$ defined  in (\ref{zdef}) converts (\ref{modeqflat}) to Bessel's equation with imaginary index $\pm i\g$, so that the solutions
may be expressed in terms of $J_{\pm i\g} (z)$. The particular linear combination in terms of a Hankel function
\be
f_{k\g(+)}(u) \equiv \sdfrac{1}{2} \sqrt{\frac{\pi}{H}}\, e^{- \frac{\pi \g}{2}} e^{\frac{i \pi}{4}}
H^{(1)}_{i \g} (z) =  \sqrt{\frac{\pi}{H}} \,\frac{e^{\frac{\pi \g}{2}}e^{\frac{i \pi}{4}}}{e^{2 \pi \g} - 1}
\,\Big[ e^{\pi \g} J_{i \g}(z) - J_{-i\g}(z)\Big]
\label{BDflat}
\ee
which has been normalized according to (\ref{Wron}), has the asymptotic behavior
\be
f_{k\g(+)}(u) \rightarrow  \frac{e^{iz}}{\sqrt{2Hz}} \leftrightarrow \frac{1}{\sqrt{2H\w_{k\g}}} \exp\big\{\!\!-i\Th_{\g}(z)\big\}
\qquad {\rm as} \qquad z \rightarrow \infty
\ee
matching the positive frequency adiabatic mode $f_k^{(0)}$ in this limit. Thus the solution (\ref{BDflat}) defines the $\vert 0,in\rag$ vacuum state
as $u \rightarrow -\infty$ according to the $m^2-i \e$ prescription, in the flat spatial sections of de Sitter space (\ref{dSflat}).
The particular solution (\ref{BDflat}) is also that of the Bunch-Davies state which is $O(4,1)$ de Sitter invariant~\cite{BunDav}.

On the other hand the particular solution to (\ref{modeqflat})
\be
f_{k\g}^{(+)} (u) \equiv \frac{\G(1 + i \g)}{\sqrt{2H\g}}\,2^{i\g} \,J_{i \g}(z) \rightarrow  \frac{z^{i\g}}{\sqrt{2H\g}} \leftrightarrow
\frac{1}{\sqrt{2H\w_{k\g}}} \exp\big\{\!\!-i\Theta_{\g}(z)\big\} \qquad {\rm as} \qquad z\rightarrow 0
\label{outflat}
\ee
is the properly normalized $|0,out\rag$ adiabatic vacuum positive frequency solution which agrees with the adiabatic form at late times,
$u \rightarrow +\infty$, in accordance with (\ref{ThetadS}). Since $f_{k\g}^{(+)} (u)$ differs from $f_{k\g(+)} (u)$, the {\it in} and {\it out}
vacuum states defined by these positive frequency solutions differ according the Feynman definition. Comparison of (\ref{outflat}) 
with (\ref{BDflat}) allows us to read off the exact Bogoliubov coefficients of the scattering problem (\ref{asymBog})
\bes
\bea
&&A_{\g} =   \frac{\sqrt{2\pi\g}\ e^{\frac{i \pi}{4}}}{2^{i\g}\,\G(1 + i \g)}
\ \frac{e^{ \frac{3\pi \g}{2}}}{e^{2 \pi \g} - 1} \\
&&B_{\g} = -\frac{\sqrt{2\pi\g}\ e^{\frac{i \pi}{4}}}{2^{i\g}\,\G(1 + i \g)}
\ \frac{e^{\frac{\pi \g}{2}}}{e^{2 \pi \g} - 1}
\eea
\label{Bogflat}\ees
which are independent of $k$ and satisfy $|A_{\g}|^2 - |B_{\g}|^2 = 1$. The square of the latter coefficient
\be
|B_{\g}|^2 = \frac{1}{e^{2\pi \g}-1} =  e^{-2\pi \g} \sum_{n=0}^{\infty} e^{-2\pi n \g} \neq 0
\label{deSBksq}
\ee
is the average number of particles created in any $\bf k$ mode at late times in the CTBD state as reckoned by the adiabatic $|0,out\rag$ vacuum.
These exact results tell us that the de Sitter invariant CTBD $|0,in\rag$ state is not the vacuum state at late times, and is unstable to pair creation,
with the average number of particles created at late times given by (\ref{deSBksq}).

Moreover from (\ref{ThetadS}) the particle creation event takes place at $z \sim \g$, at which the adiabatic phase (\ref{ThetadS}) transitions from
its large $z$ (early time) to its small $z$ (late time) behavior. Applying the complex adiabatic phase method, first using the $z$ variable,
reveals again just two complex critical points where the frequency function $\w_{k\g}^2$ vanishes, namely at
\be
z = \pm i \g
\label{zg}
\ee
analogous to (\ref{lamzeroes}) in the $E$-field case.  Evaluating (\ref{ThetadS}) at the complex critical point $-i\g$ gives
\vspace{-1mm}
\be
{\rm Im}\, \Theta_{\g}(-i \g) = \frac{\pi \g}{2}
\label{Imflat}
\ee
defining the anti-Stokes lines, and
\be
{\rm Re}\, \Theta_{\g}(z) = {\rm Re}\, \Theta_{\g}(-i\g) = 0
\label{StokesdS}
\ee
defining the Stokes lines shown the Fig. \ref{Fig:Stokes-dS-z}.

\begin{figure}[htp]
\vspace{-1cm}
\includegraphics[height=6cm,viewport= 50 50 760 600, clip]{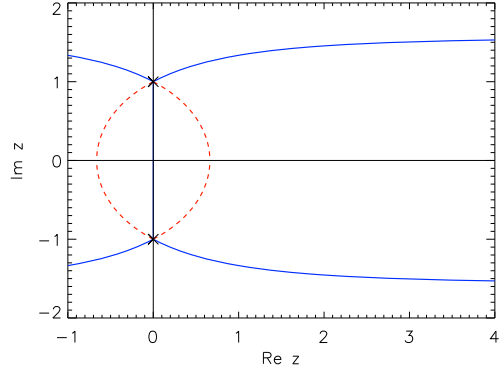}
\vspace{-7mm}
\caption{The solid (blue) lines denote the anti-Stokes lines of constant Im $\Theta_\l$ emerging from the two complex turning
points of \eqref{zg} with $\g = 1$. The dashed (red) lines are the Stokes lines of constant Re $\Theta_\l$ connecting the two critical
points, the rightmost of which crosses the real axis at $z =\ka\g= \ka$ given by (\ref{kapdef}).}
\vspace{-2mm}
\label{Fig:Stokes-dS-z}
\end{figure}

We see from (\ref{ThetadS}), (\ref{StokesdS}) and Fig. \ref{Fig:Stokes-dS-z} that the Stokes line crosses the real $z$ axis at $z=\ka\g$ or
\vspace{-1mm}
\be
u_{k\g}= H t_{\rm event}(k)= \ln \left(\sdfrac{k}{\ka\g}\right)
\label{ukgamflat}
\ee
with $\ka$ given by (\ref{kapdef}). This defines the time at which the given $k$ mode experiences its creation event from the global analysis 
of the Stokes line of the complex adiabatic phase integral crossing the real axis differs slightly from the time when the local adiabatic condition 
is maximally violated.

\begin{figure}[htp]
\vspace{-7mm}
\includegraphics[height=6cm,viewport= 50 50 760 600, clip]{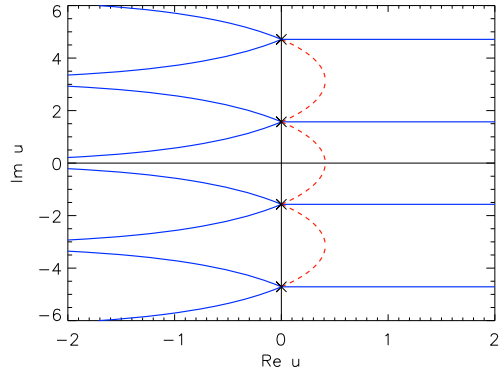}
\vspace{-7mm}
\caption{The Stokes and anti-Stokes lines of Fig. \ref{Fig:Stokes-dS-z} mapped to the complex $u$ plane, for $k=\g =1$, resulting in an infinite
number of complex critical points at $u = i\pi \left( n + \tfrac{1}{2}\right)$ along the imaginary axis, four of which are shown. The solid (blue) lines 
are the anti-Stokes lines, and the dashed (red) lines are the Stokes lines, one of which crosses the real axis at $u_{k\g} = \ln (1/\ka) = 0.411368$, 
from (\ref{ukgamflat}) with $k=\g=1$.}
\vspace{-3mm}
\label{Fig:Stokes-dS_u}
\end{figure}

The WKB adiabatic phase also determines the approximate magnitude of particle creation through
\vspace{-5mm}
\be
\vert B_{\g}\vert^2 \simeq \exp\big[\!-4\, {\rm Im}\, \Theta_{\g}(-i \g)\big] = e^{-2\pi \g}
\label{Bkgflat}
\ee
which agrees with (\ref{deSBksq}) calculated from the exact Bessel function solutions of (\ref{modeqflat}) only to leading order in $e^{-2\pi \g}$
when $\g \gg 1$. The reason for this discrepancy (and difference with the exact $E$-field result) is that the $z$ variable only spans the domain $(0, \infty)$,
unlike the infinite range $(-\infty, \infty)$ of the $u$ time variable in the $E$-field case, so that the complex turning point method utilized
previously is not strictly valid in the $z$ variable. On the other hand, if one uses the original $u=Ht$ variable of (\ref{dSflat}), which does run
over the infinite time domain, then there are an infinite number of complex turning points, at $u = \ln(k/\g) + i\pi \left( n + \tfrac{1}{2}\right), n \in \mathbb{Z}$,
four of which are shown in Fig. \ref{Fig:Stokes-dS_u}. The leading order WKB value (\ref{Bkgflat}) is the contribution from the complex turning point
in the upper half $u$ plane nearest to the real axis, which dominates if $\g \gg 1$. The infinite series of complex turning points further the from
real axis implies that there are a sum of exponentially smaller contributions in $\g$ from these additional complex turning points, and this is manifest in the
exact result (\ref{deSBksq}).

The time-dependent adiabatic particle number is defined by eq.~(\ref{adbpart}) \cite{CKHM,QVlas,HabMolMot,ShortDistDecohere,AndMot1}
with $f_k = f_{k\g(+)}$ of (\ref{BDflat}) for the initial CTBD vacuum, and where to lowest order
\bes
\bea
&&W_k^{(0)} =H\w_{k\g} =  H\sqrt{ k^2e^{-2 u} + \gamma^2}\\
&&V_k^{(1)}= -\frac{\dot \w_{k\g}}{\w_{k\g}} = H\left(1 - \sdfrac{\g^2\,}{\w_{k\g}^2}\right)
\label{Vk1}
\eea
\label{Wk0Vk1}\ees
while to second order in the adiabatic expansion
\be
W_k^{(2)} = H\left(\w_{k\g} + \sdfrac{3}{8} \sdfrac{\dot \w_{k\g}^2}{\w_{k\g}^3} -\sdfrac{1}{4} \sdfrac{\ddot \w_{k\g}}{\w_{k\g}^2} \right)
= H \w_{k\g}\, (1 + \d_{k\g}) =H \w_{k\g} + \sdfrac{H}{8\,\w_{k\g}}\left(1 - \sdfrac{\g^2\,}{\w_{k\g}^2}\right) \left(1-\sdfrac{5\g^2\,}{\w_{k\g}^2}\right) \\
\label{Wk2Vk1}
\ee
with $V_k= V_k^{(1)}$ still given by (\ref{Vk1}). A comparison of ${\cal N}^{(n)}_k(u)$ defined by (\ref{adbpart}) for both choices $n=1,2$,
along with the superadiabatic particle number defined in this case by \cite{Ber,DabDun}
\be
\overline {\cal N}_\g(u) = \frac{ |B_{\g}|^2}{4}\left\{ {\rm erfc}\! \left[\frac{-\,\Theta_\g(u)}{\sqrt{{\rm Im}\, \Theta_{\g}(-i\g)}}\right]\right\}^2
 =  \frac{1}{4\, (e^{2\pi \g}-1)} \left\{ {\rm erfc}\! \left[-\sqrt{\sdfrac{2}{\pi \g}}\, \Theta_\g(u)\right]\right\}^2
\label{superadbdS}
\ee
normalized to the correct value of $|B_\g|^2$ in (\ref{deSBksq}) is shown in Fig. \ref{Fig:beta2dS}. This confirms that the particle number rises
rapidly as the Stokes' line is crossed, the global analysis of the adiabatic phase in the complex plane determining most accurately the time
of the particle creation event (\ref{ukgamflat}) \cite{AndMot1,DabDun}.

\begin{figure}[htp]
\includegraphics[height=6cm,viewport= 0 0 680 480, clip]{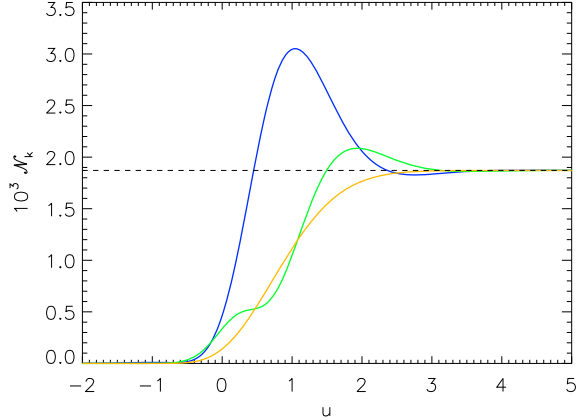}
\vspace{-7mm}
\caption{The mean number of particles created from the vacuum as a function of time in de Sitter space for $\g=1$ and $k=1$.  The three curves are
for adiabatic particle numbers ${\cal N}_k^{(j)}$ defined by different orders of the asymptotic expansion (\ref{asymexp}) and (\ref{Wk2Vk1})\cite{AndMot1},
for $j=1,2$, and the `superadiabatic' particle number defined by (\ref{superadbdS}) \cite{DabDun}. Note that the zeroth order adiabatic curve (blue) has 
the highest peak while the superadiabatic curve (orange) has no peak.  Although differing somewhat in transient details around $u=u_{k\g} =0.411368$
of (\ref{ukgamflat}), all three curves rise rapidly from zero near there and tend to the same asymptotic value $|B_{\g }|^2 = 1.87094 \times 10^{-3}$ of 
(\ref{deSBksq}) as $u\rightarrow \infty$.}
\vspace{-5mm}
\label{Fig:beta2dS}
\end{figure}

The vacuum decay rate for the expanding half of de Sitter space covered by the Poincar\'e flat spatial coordinates (\ref{dSflat})
starting in the $|0, in \rag$ CTBD vacuum can be determined by the Differential Method (\ref{DecayRateDiff}). In this case 
only the magnitude $|{\bf k}| = Hk$ is fixed by the Stokes line crossing, so that inverting (\ref{ukgamflat})
\vspace{-5mm}
\be
\overline k (t) = \ka \g\, e^{Ht}= \ka \g\, a_{\rm dS}(t)
\label{overk}
\ee
gives the value of $k$ of the mode experiencing its creation even at time $t$. Since the integration measure in (\ref{DecayRateDiff}) is
$d^3{\bf k} = H^3\, k^2 dk\, d\W_{\bf \hat k}$ and the four-volume factor is $d{\cal V}_4 = V a_{\rm dS}^3(t) dt$, we have from (\ref{overk}) the Jacobian
\be
\frac{d^3 {\bf k}}{d {\cal V}_4} \bigg\vert_{ k = \overline{k} (t)} = \frac{H^3\,d\W_{\bf \hat k}}{V a_{\rm dS}^3}\, \frac{\overline k^2 d\overline k}{dt}
= \frac{\ka^3 \g^3H^4\!}{V}\  d\W_{\bf \hat k}
\label{dSJacob}
\ee
which is independent of $t$, the factors of $a_{\rm dS}^3(t)$ having cancelled. Thus only the integral over the directions of $\bf \hat k$
remains in (\ref{DecayRateDiff}), which since $\int d\W_{\bf \hat k} = 4 \pi$ and $N=1$ for a single real scalar, gives
\be
\G_{\rm dS} = \frac{\ka^3\g^3H^4}{4\pi^2}\, \ln \big(1 + |B_{\g}|^2\big) = -\frac{\ka^3 \g^3H^4}{4 \pi^2 }\, \ln \left(1- e^{-2 \pi \g}\right)
\label{DecayBD}
\ee
for the decay rate of the CTBD `vacuum' state of de Sitter space into particle pairs of mass $m = H \sqrt{\g^2 + \frac{1}{4}}$,
with $\ka$ given by (\ref{kapdef}). The decay rate (\ref{DecayBD}) of the de Sitter invariant Bunch-Davies `vacuum,' determining
the finite pre-factor by the Stokes line crossing is a principal result of our analysis. Interestingly $\G_{\rm dS}$ tends to zero in the limit
$\g \rightarrow 0$ as $- \g^3 \ln \g$, while
\be
\G_{\rm BD} \rightarrow \frac{\ka^3 m^3 H}{4 \pi^2}\,  \exp \left(- \sdfrac{2 \pi m}{H} \right) \qquad {\rm for} \quad m \gg H
\label{RateBDlargem}
\ee
in the large mass or flat space limit, similarly to (\ref{vacdecayE}) for the electric field case.  We note from (\ref{overk}) that the physical wavelength
$(a_{\rm dS}/H\,\overline k) = (\ka\g H)^{-1}$ of the Fourier mode at the time of its particle creation event, is of the order of the de Sitter
Hubble horizon if $\g \sim 1$, but can be much smaller than the horizon if $m\gg H, \g \gg 1$.

Note also that although there is no integral over $k$ to perform in (\ref{DecayBD}), this value of $\G_{dS}$ obtained from the Differential Rate
definition (\ref{DecayRateDiff}) is identical to what would be obtained by an integral rate formula in pure de Sitter space
if the de Sitter window step function value of
\be
|B_k|^2 = \left\{ \begin{array}{cc} |B_{\g}|^2\,, & \qquad \overline k (t_0) \le  k \le \overline k(t_1) \\ 0\,, & \qquad {\rm otherwise} \end{array}\right.
\label{puredS}
\ee
were used. Because of the kinematic factor of $k^2 dk$ in (\ref{inout}) the integral is clearly dominated by the largest value of $k$ contributing at 
the largest value of the FLRW scale factor for an expanding universe, and one may replace the lower limit of $\overline k(t_0) =\ka\g a_{\rm dS}(t_0)$
in (\ref{puredS}) by zero, in the limit of large $a_{\rm dS}(t_1) = e^{u_1}$. Thus (\ref{inout}) with (\ref{puredS}) leads again to (\ref{DecayBD}),
if divided by the integrated four-volume $V \int_{t_0}^{t_1} dt\, a_{dS}^3(t)\rightarrow \frac{1}{3}V e^{3u_1}$ in the same limit.

The result (\ref{DecayBD}) is half of what would be obtained in global de Sitter space in leading exponential order for the closed ${\mathbb S}^3$ spatial
sections in the same limit, the reason being there are two creation events in each $k$ mode in the closed spatial sections, one in the contracting
phase and one in the expanding phase. Thus except for one creation event in each mode as opposed to two, the same phenomenon of vacuum decay 
takes place in the Poincar\'e patch of a de Sitter universe that is only expanding, usually considered in FLRW cosmological models, as in the globally 
complete closed ${\mathbb S}^3$ spatial sections. The vacuum decay rate (\ref{DecayBD}) also differs from the result of \cite{MotPRD85,AndMot1} 
by a finite pre-factor because of the difference of $N=1$ {\it vs.}~$N=2$, and the differing estimate of the constant pre-factor in the $K$ cutoff of the 
mode sum in (\ref{inout}), which is determined to be $\overline k(t_1)$ in the present work by the detailed analysis of the particle creation event 
in real time by the Stokes line crossing. 

\section{Adiabatic Switching de Sitter On and Off}
\label{Sec:dSadb}

As in the $E$-field case, if the Integral Method for defining the decay rate (\ref{DecayRateInt}) were used, one would need a sharp
step `window' function cutoff for the allowed $k$ values of Fourier modes undergoing particle creation events in a finite time in order
to reproduce (\ref{DecayBD}).  Since the de Sitter phase cannot end abruptly without violating the adiabatic condition (\ref{adbcond}), 
one may ask if this assumption and the rate (\ref{DecayBD}) can be obtained also by adiabatically switching the de Sitter background 
on and off in the limit in which the time the applied de Sitter background is switched on is taken to infinity, as in Sec.~\ref{Sec:Eadb} for the
$E$-field case. To address this question we investigate two different time profiles, the first with a single adiabatic parameter
\vspace{-1mm}
\be
h(t) \equiv \frac{\dot a }{a} = H\, \sech^2(t/T)
\label{h1param}
\ee
suggested by analogy to (\ref{Esech}), and the second
\be
h(t) \equiv \frac{\dot a}{a} =  \frac{H}{2} \tanh \big[b\,(t - t_0)\big] - \frac{H}{2} \tanh \big[b\,(t - t_1)\big]
\label{h2param}
\ee
suggested by the two-parameter $(b, T)\ E$-field profile (\ref{Eprofile}) illustrated in Fig.~\ref{Fig:Eprofile}.  

\begin{figure}[hbp]
\vspace{-1mm}
\includegraphics[height=6cm,viewport= 0 0 680 480, clip]{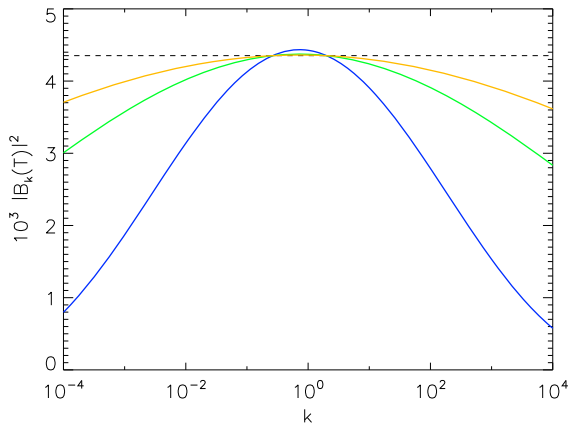}
\vspace{-7mm}
\caption{Particle creation $|B_k(T)|^2$ for the one parameter quasi-de Sitter profile (\ref{h1param}), as a function of $k$ for
$m=H$. The curve that falls off the most rapidly in $k$ (blue) is for $HT= 20$, the middle one (green) is for $HT= 40$, and
the outer one (orange) is for $HT = 60$, showing the approach to $|B_{\g}|^2 = 4.35228 \times 10^{-3}$ of Eq. (\ref{deSBksq})
for small $k/a$. Compare to Fig.~\ref{Fig:B21paramEfield}.}
\vspace{-4mm}
\label{Fig:B21paramdS}
\end{figure}

In the first case (\ref{h1param}) the FLRW scale factor may be taken to be
\vspace{-1mm}
\be
a(t|T) = \exp \left[HT \tanh(t/T)\right]
\label{a1param}
\ee
with an arbitrary multiplicative constant of integration set equal to unity. As $t \rightarrow \mp \infty$, $a(t|T)$ goes to a constant
and the flat space vacua are uniquely defined. Since the solution of the mode eq.~(\ref{harmosc}) with (\ref{FLRWomega})
is not known analytically for this scale factor, we present the numerical results for particle creation $|B_k|^2$ in Fig.~\ref{Fig:B21paramdS}, 
which may be compared to Fig.~\ref{Fig:B21paramEfield}. As in the electric field profile (\ref{Esech}), $|B_k(T)|^2$ falls off at large momenta 
for any finite $T$, the falloff becoming more and more gradual as $T$ becomes larger, in which limit a flat plateau at small $k$ characteristic 
of the constant $h\rightarrow H$ de Sitter value (\ref{deSBksq}) is attained. Again the $k \rightarrow \infty$ (fixed $T$) and $T\rightarrow \infty$
(fixed $k$) limits of $|B_k(T)|^2$ do not commute, and the gradual falloff of $|B_k(T)|^2$ for those $k$ going through their creation events 
when $h(t)$ is not constant makes the FLRW time profile (\ref{a1param}) inappropriate for the Integral Method of determining the
decay rate for pure de Sitter space. However, again as in the case of the one parameter time profile (\ref{Esech}),
the Differential Method for determining the vacuum decay rate of de Sitter space may be applied to the FLRW time profile
(\ref{a1param}) and its $|B_k(T)|^2$ in the adiabatic limit $T\rightarrow \infty$, provided the differential Jacobian (\ref{dSJacob})
is computed for the modes in the central plateau of Fig. \ref{Fig:B21paramEfield} where $h(t) \rightarrow H$ is constant
and the result for pure de Sitter space (\ref{DecayBD}) is reobtained.

In the second case of the time profile (\ref{h2param}) the FLRW scale factor may be taken to be
\be
a(t\vert t_0, t_1, b) = \exp\left\{ \frac{H}{2b}\, \ln \left[ \frac{\cosh \big[b\, (t-t_0)\big]}{\cosh \big[b\, (t-t_1)\big]}\right] + \frac{ H(t_0 + t_1)}{2}\right\}
\label{a2param}
\ee
in which the multiplicative constant of integration has been chosen so that the scale factor has the simple behaviors
\vspace{-3mm}
\be
a(t\vert t_0, t_1, b) \rightarrow \left\{ \begin{array} {cc} \ e^{Ht_0}& \qquad t \ll t_0 \\
e^{Ht}& \qquad t_0 \ll t \ll  t_1 \\
\ e^{Ht_1}&\qquad  t_1 \ll t
\end{array}\right.
\label{atwoparam}
\ee
in each region for $b(t_1 - t_0) \gg 1$. Thus, as in (\ref{a1param}), the scale factor is a constant in both the very early and very late time limits,
the spacetime becomes flat in those limits and both the $|0,in\rag$ and  $|0, out\rag$ vacuum states and the particle number are unambiguously 
well-defined for $t \rightarrow \mp \infty$. The $k$ integral in the probability overlap (\ref{inout}) again is finite. The de Sitter phase for $t_0 < t < t_1$ 
in between can be made arbitrarily long, while the adiabatic turning on and off of the de Sitter phase takes a finite time of order $b^{-1}$, which 
needs to be large enough so that the transition is gentle and adiabatic, and does not in itself lead to significant particle creation, {\it i.e.} $b \ll H$.

\begin{figure}[tbp]
\begin{center}
\vspace{-3mm}
\includegraphics[height=6cm,viewport= 0 0 680 480, clip]{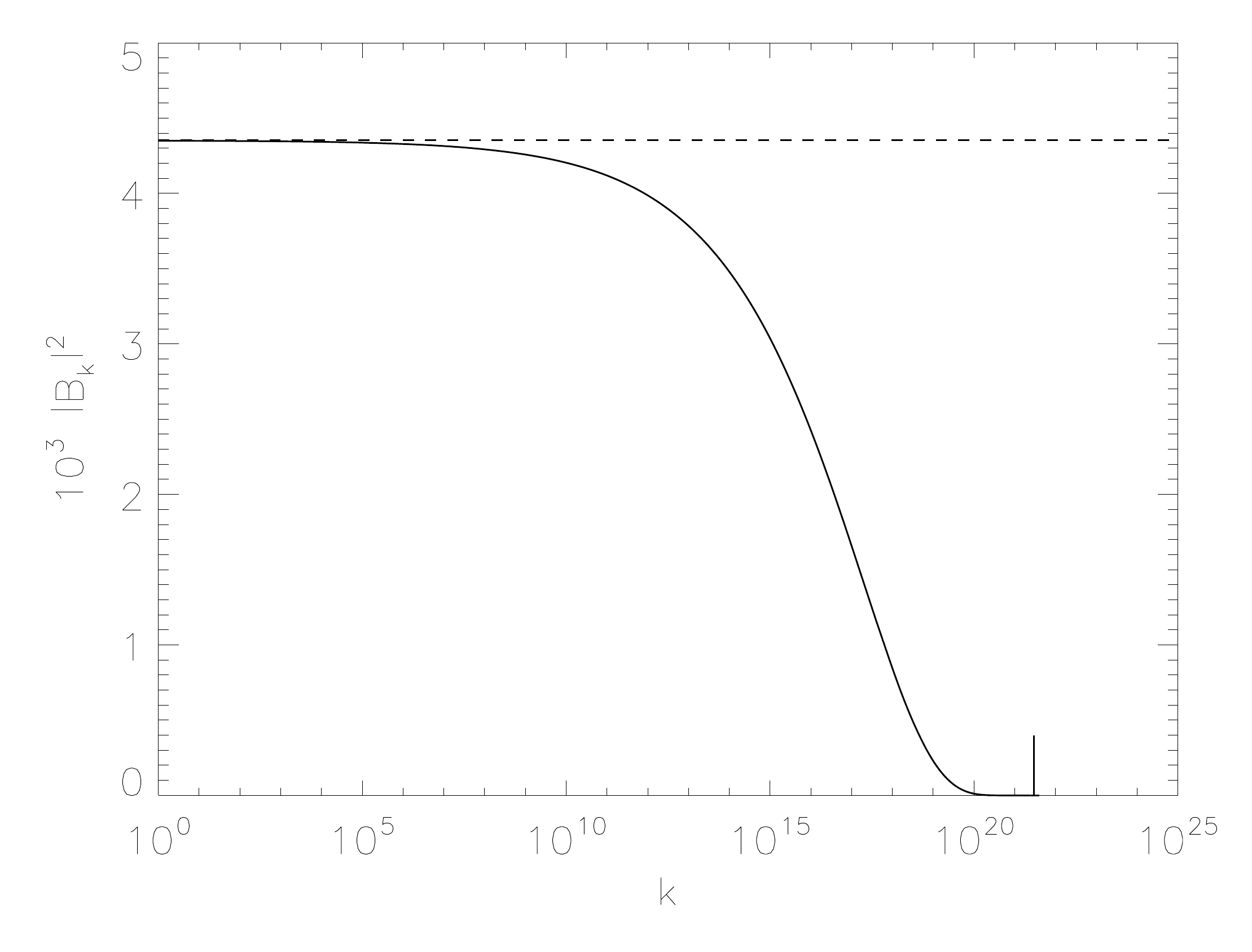}
\vspace{-6mm}
\caption{The number density of created particles $|B_k|^2$ in the final state after the de Sitter phase is switched off according to
the time profile (\ref{a2param}), for $u_1=50, b=0.1\,H$ and $m = H$. The dashed line is the constant (\ref{deSBksq}), here 
$4.35228\times 10^{-3}$ expected for pure de Sitter space and $\g= \sqrt{3}/2$, and the solid hash marker is the value
of $\overline k= \ka \g e^{u_1}= 2.97577 \times 10^{21}$ expected from (\ref{overk}).}
\label{Fig:B2}
\end{center}
\vspace{-7mm}
\end{figure}

Fig.~\ref{Fig:B2} shows numerical results for the particle number $|B_k|^2$ in the final static region, as $u\rightarrow +\infty$.
Note that the pure de Sitter value of $|B_k|^2$, (\ref{deSBksq}) is obtained for small $k \ll \g e^{u_1}$. However the falloff
from this constant de Sitter `plateau'  value is very gradual unlike the integrand in Fig. \ref{Fig:Ebetas}. The value of $|B_k|^2$ also 
begins to fall off markedly at $k$ values much smaller than the value $\ka \g e^{u_1}$ expected from (\ref{overk}) and (\ref{deSBksq}).
In the Integral Rate formula (\ref{DecayRateInt}), the integral over $dk$ is weighted by $k^2$. This integrand is shown in Fig. \ref{Fig:k2logB2},
which because of the falloff of $|B_k|^2$ at large $k$ achieves a maximum value still considerably less than would be expected
from the pure de Sitter result (\ref{deSBksq}), and at a considerably lower value of $k$ than $\ka \g e^{u_1}$.

\begin{figure}[htp]
\begin{center}
\vspace{-3mm}
\includegraphics[height=6cm,viewport= 0 0 760 500, clip]{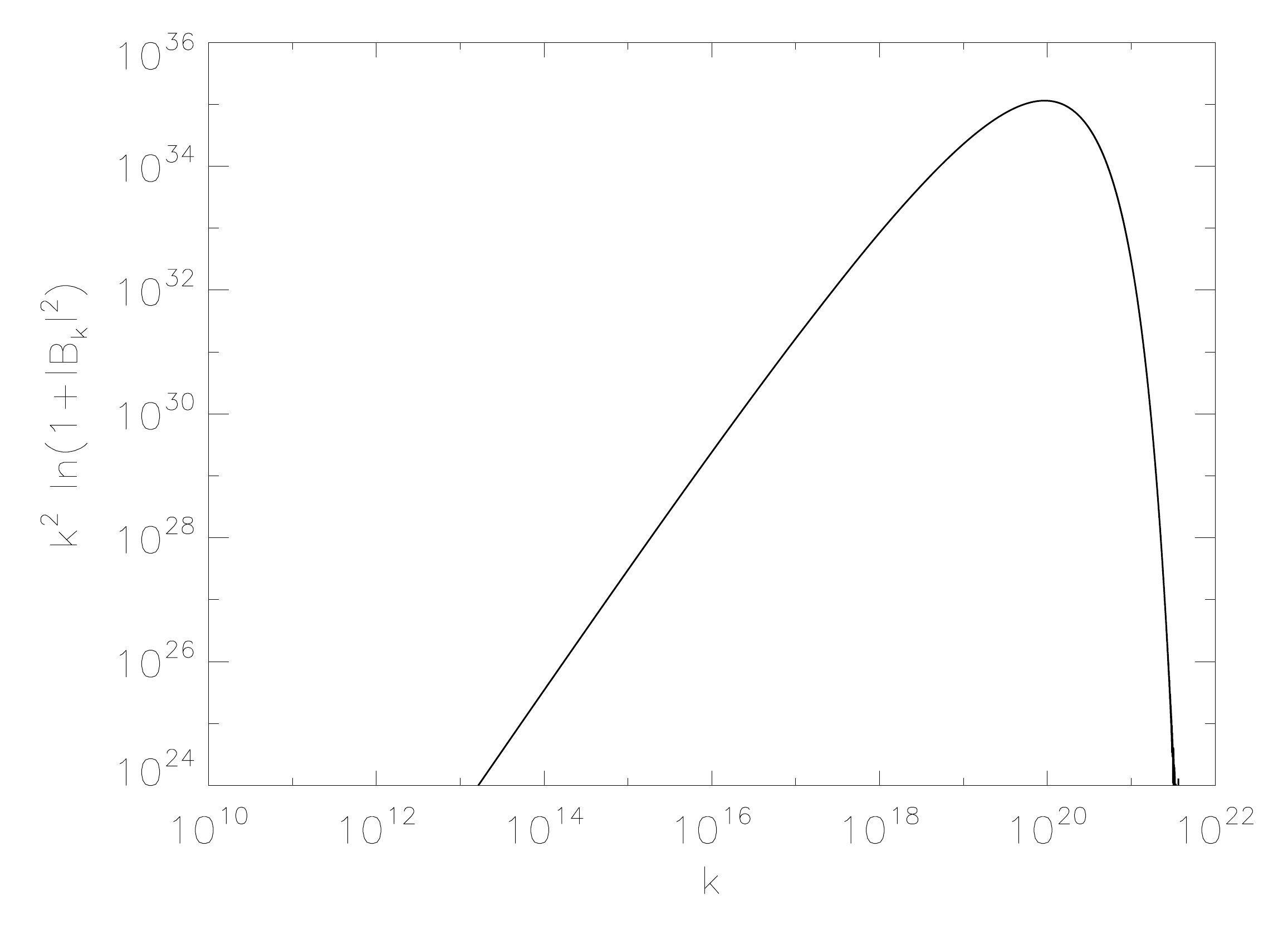}
\vspace{-6mm}
\caption{The integrand $k^2 \ln(1+|B_k|^2)$ of (\ref{DecayRateInt}) for the scale factor (\ref{a2param}) on a log-log plot
for the case $u_1=50, b=0.1\,H$ and $m = H$.}
\label{Fig:k2logB2}
\end{center}
\vspace{-8mm}
\end{figure}

In the Differential Rate calculation of (\ref{DecayBD}) the Jacobian factor (\ref{dSJacob}) is time independent, so that in the Integral Rate
the $u_1$ volume dependence in the integrated four-volume
\be
{\cal V}_4 = V \int_{t_0}^{t_1} dt\, a^3 (t\vert t_0, t_1, b) \rightarrow \frac{V a^3 (t_2\vert t_1, t_2, b)}{3} \rightarrow  \frac{V}{3} \, e^{3 u_1}\, e^{-\frac{H}{2b}\ln 2}
\label{fourvol}
\ee
should be cancelled by the range of $k$ integration in the integral $\int k^2 dk \ln(1 + |B_k|^2)$ for large $u_1=Ht_1$. Fig. \ref{Fig:Scaling} show the 
independence of the value of the plateau value of $|B_k|^2$ for small argument $k e^{-u_1} \rightarrow 0$, hence large $u_1$. This shows that 
the plateau value of $|B_k|^2$ exists for small enough $ke^{-u_1}$, and the scaling expected and needed for the dependence on the final time $u_1$ 
to drop out of the rate. 

\begin{figure}[htp]
\vspace{-9mm}
\includegraphics[height=6cm,viewport= 0 0 760 500, clip]{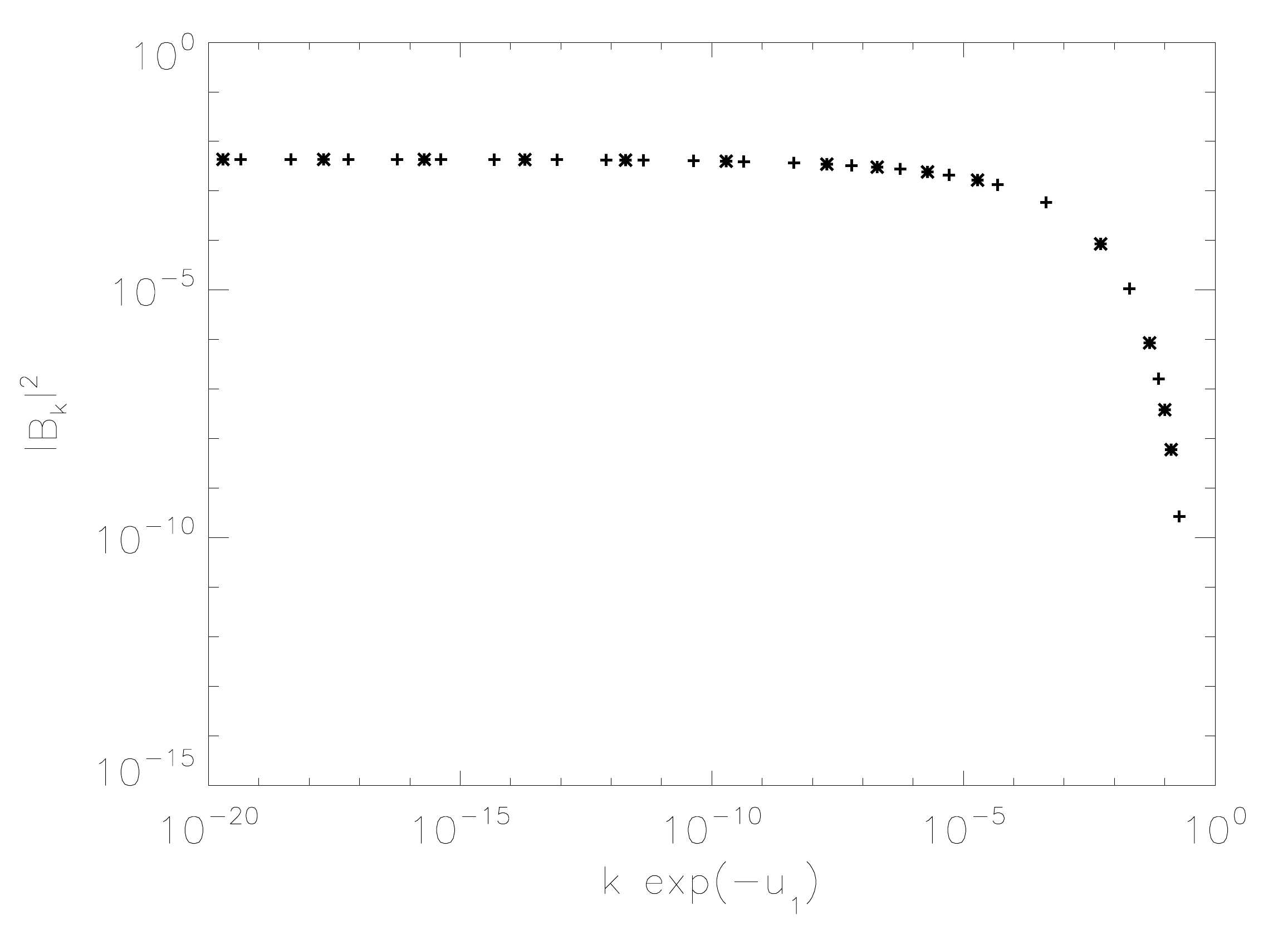}
\vspace{-7mm}
\caption{The mean particle number $|B_k|^2$ as function of rescaled $k e^{-u_1}$ for $u_1 =50$ (stars) and $u_1=70$
(crosses) in the case $b=0.1\,H, m = H$, showing its universal scaling behavior at large $u_1$.}
\vspace{-2mm}
\label{Fig:Scaling}
\end{figure}

However, if we try to apply the Integral Method (\ref{DecayRateInt}) to define the vacuum decay rate by means of the profile (\ref{a2param}), 
the long gradual tail in $|B_k|^2$ as a function of $k$ for larger $k$, yields a rate that depends on $b$ no matter how large $u_1$ is.
In Fig.~\ref{Fig:B2b} we show the dependence of $|B_k|^2$ as a function of rescaled $ke^{-u_1}$ for various values of $b$, showing that the
falloff from its de Sitter plateau value depends on $b$, and occurs at a smaller value of $ke^{-u_1}$ for smaller values of $b$. This implies
a smaller contribution to the integral $\int k^2 dk \ln(1 + |B_k|^2)$ for smaller $b$.

\begin{figure}[htp]
\vspace{-3mm}
\includegraphics[height=6cm,viewport= 0 0 760 500, clip]{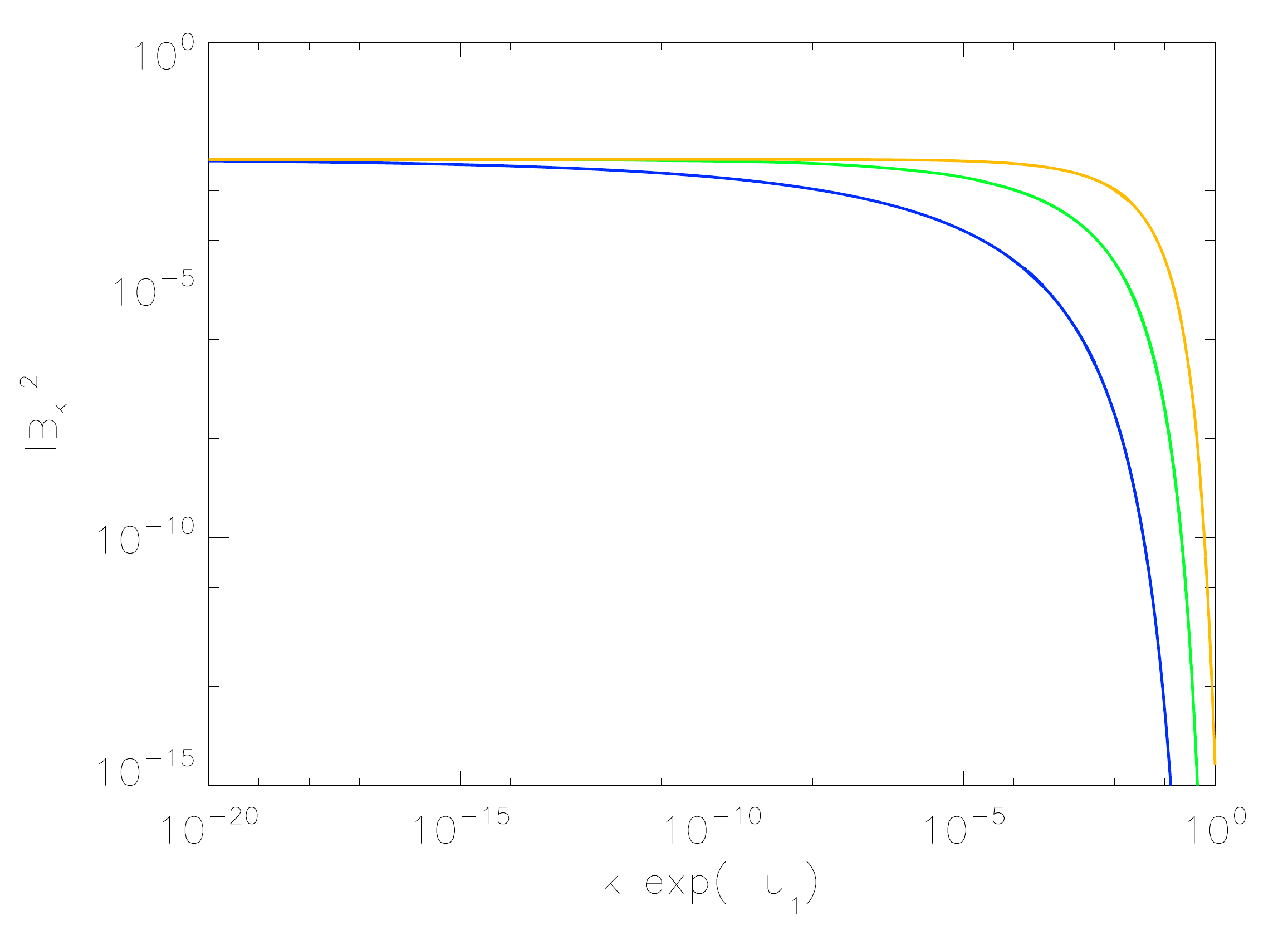}
\vspace{-5mm}
\caption{Mean particle number $|B_k|^2$ {\it vs.} rescaled momentum $k e^{-u_1}$ for the profile (\ref{a2param}) for the cases lower to upper of $b = 0.05\,H$ (blue), 
$b = 0.1\,H$ (green), and $b = 0.2\,H$ (orange). For the range of values shown the data for the curves was computed for values of $u_1$ that are in 
the scaling range where to a good approximation the value of $|B_k|^2$ depends only on the product $k e^{-u_1}$. }
\vspace{-4mm}
\label{Fig:B2b}
\end{figure}

Indeed Fig.~\ref{Fig:Rateb} shows the numerical results for the decay rate (\ref{DecayRateInt}) turning de Sitter space on and off according to the profile 
(\ref{a2param}). As expected, the decay rate rises for large $b$ due to breakdown of the adiabatic condition (\ref{adbcond}) and the creation of 
particles during the switching on and off of the de Sitter phase in the short time $b^{-1}$, coming to dominate over the particle creation in the de Sitter 
phase itself, so we should exclude these large values of $b$. As $b$ is decreased the rate decreases due to the more rapid falloff of the integrand
shown in Fig.~\ref{Fig:B2b}, reaching a minimum value of (\ref{DecayRateInt}) at $b\simeq 0.1\,H$, with a rise again for smaller $b$. This rise
for smaller $b$ is the result of the multiplicative exponential dependence of ${\cal V}_4$ in (\ref{fourvol}), rather than additively as in the $E$-field case.

\begin{figure}[htp]
\vspace{-7mm}
\includegraphics[height=6cm,viewport= 50 50 760 600, clip]{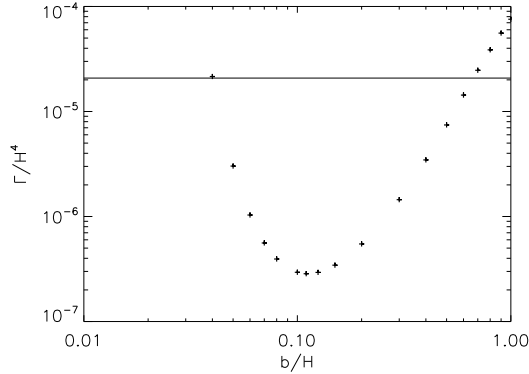}
\vspace{-7mm}
\caption{Numerical results for the decay rate calculated by the Integral Method (\ref{DecayRateInt}) for the scale factor profile
(\ref{a2param}) and integrated four-volume (\ref{fourvol}) as a function of $b$ is shown on a log-log plot for $m = H, \g = \sqrt{3}/2$. The horizontal
line is the predicted pure de Sitter rate (\ref{DecayBD}) of $2.079895 \times 10^{-5}$ for this value of $\g$.}
\vspace{-3mm}
\label{Fig:Rateb}
\end{figure}

The minimum in $b$ shows that there is de Sitter vacuum decay, no matter how slowly or rapidly the de Sitter phase is ended, but that the 
probability of decay in the profile (\ref{a2param}) depends upon the time and manner spent exiting the de Sitter phase around $t \sim t_1$,
and the $b$ dependence never drops out entirely. This makes the profile (\ref{a2param}) finally inappropriate for attempting to determine the 
the de Sitter vacuum decay rate by the Integral Method (\ref{DecayRateInt}). Indeed the form of $|B_k|^2$ for large $k$ in Figs. \ref{Fig:B2}-\ref{Fig:Scaling} 
with its gradual fall-off from the de Sitter plateau value for the FLRW profile (\ref{a2param})-(\ref{atwoparam}) actually is more similar to
that obtained in the {\it single parameter} profiles (\ref{Esech}) or (\ref{a1param}), in which the deviation from the constant plateau value
characteristic of the persistent $E$-field or de Sitter background cannot be eliminated no matter how large $T$ or $u_1$ is made. In those 
cases we observed that we could only use this profile with a time varying electric field and its Bogoliubov coefficient (\ref{BkadbT}) to recover 
the Schwinger rate appropriate for a constant $E$-field from the {\it Differential} rate formula (\ref{DecayRateDiff}), and then only by taking the 
$T\rightarrow \infty$ limit. The pure de Sitter rate can likewise be extracted from the two-parameter profile only in the differential rate formalism of
Sec.~\ref{Sec:dSPersist}. This reproduces (\ref{DecayBD}), since by its scaling behavior, {\it cf.} Fig \ref{Fig:Scaling}, for large $u_1$ the value of $|B_k|^2$ 
approaches its small $k$ de Sitter plateau value $|B_{\g}|^2$ of (\ref{deSBksq}), independent of $b$ and analogous to the $T\rightarrow \infty$ 
limit of (\ref{Bsqlim}). Since in this constant $H$ de Sitter regime the differential formula (\ref{Bsqlim}) holds, the result of applying the Differential 
Method (\ref{DecayRateDiff}) is again (\ref{DecayBD}), with no integral over $k$ necessary.

The failure of the FLRW trial profile (\ref{a2param}) to reproduce the de Sitter rate (\ref{DecayBD}) by the Integral Rate formula (\ref{DecayRateInt})
for arbitrarily large $u_1$ is nevertheless interesting, and stands in marked contrast to the corresponding calculation with the $E$-field profile
(\ref{Eprofile})-(\ref{Aprofile}) with the Schwinger rate recovered in the extrapolation to the limit of of large $T$ in Fig.~\ref{Fig:ErateT}.
It shows that there is greater sensitivity to the switching off of the de Sitter background simultaneously over an exponentially large volume
at late times in the expansion, reflected both in the dependence upon $b$ of the tail of the particle distribution going through their creation events
as the de Sitter phase ends shown in Fig.~{\ref{Fig:B2b} and the multiplicative exponential dependence on $b$ of the four-volume ${\cal V}_4$ in 
(\ref{fourvol}). This greater sensitivity to how the de Sitter phase is ended at distances much greater than the de Sitter future event horizon perhaps is
unsurprising, since it is quite unlike flat spacetime in which turning off a uniform electric field everywhere simultaneously in Minkowski time presents 
no obstacle to causality. In the latter case one recovers the Lorentz invariant Schwinger rate in the limit of large $T$, while the failure to recover
the de Sitter vacuum decay rate (\ref{DecayBD}) in the former case suggests that not only de Sitter invariance is necessarily broken, no matter how
long the de Sitter phase lasts, but that spatial homogeneity may also be spontaneously broken, in that there is a residual sensitivity to an infrared 
spatial cutoff at the horizon scale, necessary to restrict any spatiotemporal variation of $H$ to within a single causal Hubble horizon. 

\section{Energy and Pressure of Created Particles: Backreaction}
\label{Sec:Ener}

The results of the previous sections indicate that so long as the exit from the de Sitter phase is gentle enough, any particles created
during that phase end up as particles in the asymptotically static region where the definition of a particle is unambiguous. This shows that 
the adiabatic particle definition of ${\cal N}_k^{(n)}$ for either $n=1,2$ used in (\ref{Wk0Vk1}) or (\ref{Wk2Vk1}) during the de Sitter phase \cite{AndMot1}
is robust and survives in the final asymptotic flat space region as $|B_{\bf k}|^2$. There is no doubt that these are the real particles observed in the 
final state after the time-dependent background has been turned off. This may be verified also by evaluating the energy density and pressure
of the created particles. After subtracting the vacuum value of the stress tensor components obtained by setting $A_{\bf k}=1, B_{\bf k} =0$,
we obtain for the renormalized flat space energy density simply \cite{HabMolMot,ShortDistDecohere}
\be
\varepsilon =  \lag T_{tt} \rag_{\!_R} = \frac{1}{a^3} \int \frac {d^3{\bf k}\,}{(2\pi)^3} \ \w_{\bf k} \, |B_{\bf k}|^2
\label{ener}
\ee
where here $\w_{\bf k}$ and $a\rightarrow e^{u_1}$ are the constant values of the frequency (\ref{FLRWomega}) and scale factor
in the asymptotic late time for the profile (\ref{a2param})-(\ref{atwoparam}) after the expansion has been turned off and space is again flat.
The corresponding expression for the renormalized isotropic pressure in flat space is
\be
p = \frac{1}{3a^3}\! \int\!\frac{\!\!d^3{\bf k}\!}{(2\pi)^3} \left(\w_{\bf k} - \frac{m^2}{\w_{\bf k}}\right)|B_{\bf k}|^2
+ \frac{1}{3a^3}\! \int\!\frac{\!\!d^3{\bf k}\!}{(2\pi)^3\!}\left[2\,(6\xi-1)\,\w_{\bf k}- \frac{m^2}{\w_{\bf k}}\right]{\rm Re} \left(A_{\bf k} B_{\bf k}^* e^{-2i\w_{\bf k} t}\right)
\label{press}
\ee
where $6\xi -1 = 0$ in the present study. Note that because of the scaling behavior of $|B_k|^2$ illustrated
in Fig. \ref{Fig:Scaling}, the change of variable from $\bf k$ to ${\bf k}/a$ shows that both the energy density and the pressure
are {\it constants, independent} of the length of time $u_1$ spent in the de Sitter phase. In other words $\varepsilon$ and $p$
{\it do not redshift to zero} with the exponential de Sitter expansion. The reason for this is that although each $k$ mode
certainly does redshift with the expansion, particles are continually being created by the expansion at the latest time
$u_1$ to replenish them at the largest $k \sim \overline k(t_1) =\ka\g e^{u_1}$, so that the integrals (\ref{ener}) and (\ref{press}) are
independent of $u_1$.

\begin{figure}
\vspace{-5mm}
\hspace{-1cm}
\includegraphics[height=6cm,viewport= 0 0 680 490, clip]{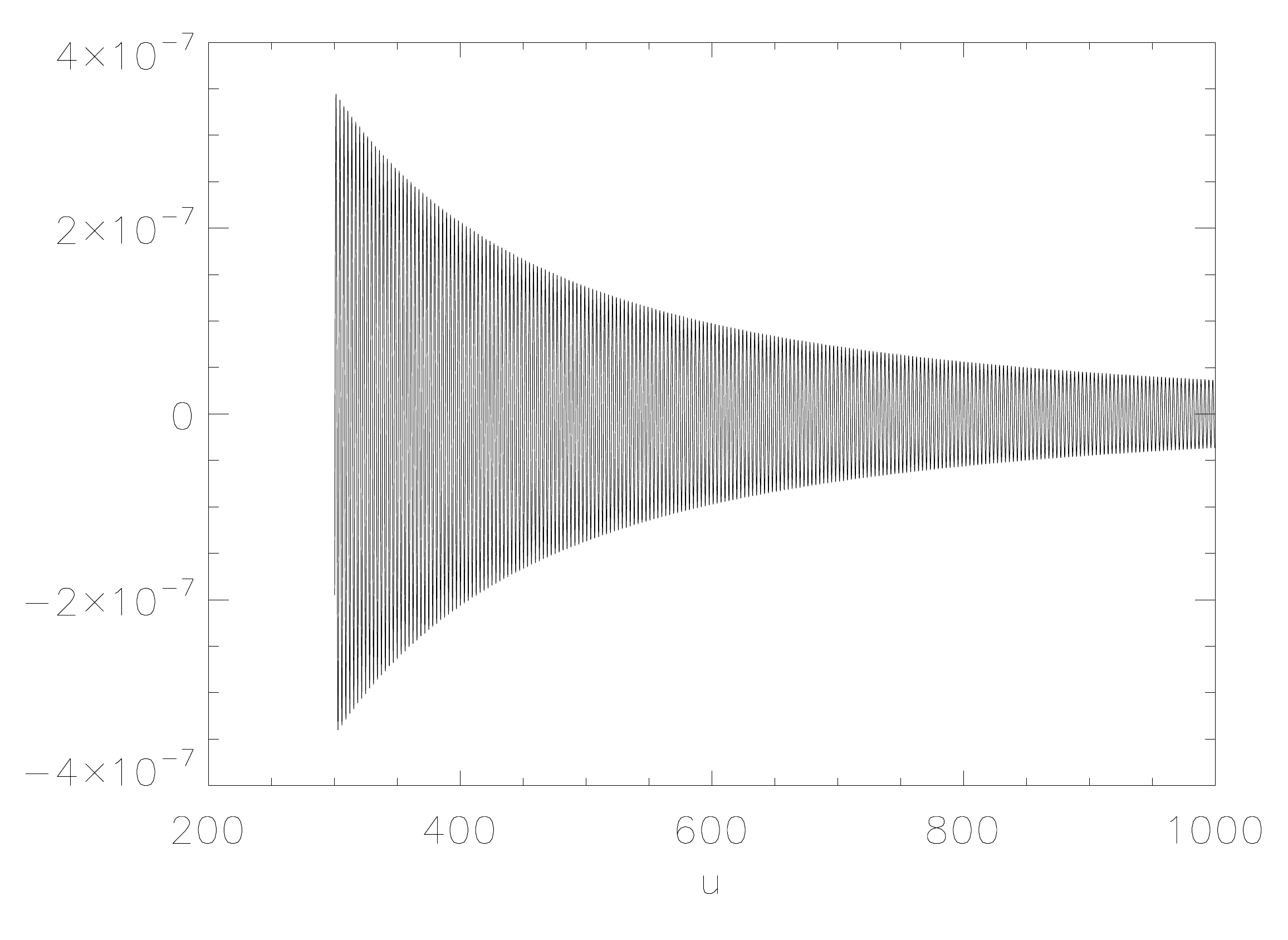}
\vspace{-7mm}
\caption{The phase coherent oscillating part of the pressure, given by the last term in (\ref{press}), for $\xi = \frac{1}{6}, m = H,u_1 = 10$,
and $b = H$. The envelope of the rapid oscillations falls as $1/u^2$ at late times.}
\label{Fig:posc}
\vspace{-4mm}
\end{figure}

If these integrals are evaluated for the pure de Sitter window value (\ref{puredS}) and large $u_1$, then making the change
of variable~\eqref{zdef} we obtain
\bes
\bea
\varepsilon &=& \frac{H^4\, |B_{\g}|^2}{2\, \pi^2}  \int_0^{\ka\g} dz\, z^2\left(z^2 + \sdfrac{m^2}{H^2}\right)^{\!\frac{1}{2}}\nn
&=& \frac{H^4\, |B_{\g}|^2}{16\, \pi^2}
\left\{ z \left(2z^2 + \sdfrac{m^2}{H^2}\right)\left(z^2 + \sdfrac{m^2}{H^2}\right)^{\!\frac{1}{2}}
- \sdfrac{m^4}{H^4} \ln\left( \frac{H}{m} \left[z + \left(z^2 + \sdfrac{m^2}{H^2}\right)^{\!\frac{1}{2}}\right]\right)\right\}_{z=\ka\g}\\
p &=& \frac{H^4\, |B_{\g}|^2}{6\, \pi^2}    \int_0^{\ka\g} dz\, z^4\left(z^2 + \sdfrac{m^2}{H^2}\right)^{\!-\frac{1}{2}}\nn
&=& \frac{H^4\, |B_{\g}|^2}{16\, \pi^2}
\left\{ z \left(\sdfrac{2}{3}z^2 -\sdfrac{m^2}{H^2}\right)\left(z^2 + \sdfrac{m^2}{H^2}\right)^{\!\frac{1}{2}}
+ \sdfrac{m^4}{H^4} \ln \left( \frac{H}{m}\left[z + \left(z^2 + \sdfrac{m^2}{H^2}\right)^{\!\frac{1}{2}}\right]\right)\right\}_{z=\ka\g}
\eea
\label{enerpress}\ees
where $|B_{\g}|^2$ is given by (\ref{deSBksq}), $\ka$ is given by (\ref{kapdef}), we have taken $\xi= \frac{1}{6}$ and
also neglected the last interference term in (\ref{press}). This is justified because as shown in Fig.~\ref{Fig:posc},
this term oscillates rapidly in the static out region and vanishes in the late time limit, just as the oscillating $A_{\bf k}B^*_{\bf k}$
quantum interference term in the electric current does at late times illustrated in the Left Panel of Fig.~\ref{Fig:jzoscT}.
Because of the rapid oscillations and their damping in evidence in Fig. \ref{Fig:posc}, there is very effective phase decoherence 
or {\it `dephasing'} in these terms, and the contribution of the interference term in the mean pressure washes out. This behavior 
is related to the fact that $|B_{\bf k}|^2$ and the diagonal elements of the density matrix in the particle basis (\ref{rhopair}) or (\ref{rhopairch}) 
are adiabatic invariants, whereas the $A_kB_k^*$ interference terms and off-diagonal elements of the density matrix depend upon the phase 
$\exp(-2i \Theta_{\bf k})$, which rapidly oscillates as a function of either $t$ or $k$ in flat space. Thus, if the rapidly oscillating
off-diagonal element of the density matrix in the final state basis are neglected, the initial pure vacuum state $|0, in\rag$ 
may be treated as a mixed state with positive entropy in the late time {\it out} basis, and the particle creation, in principle 
unitary and reversible if all exact phase correlations are preserved, becomes effectively Markovian and irreversible \cite{CKHM}.

The values of the energy and pressure in the asymptotic final state are independent of the duration of the de Sitter phase
because of the scaling illustrated in Fig. \ref{Fig:Scaling}, and both are {\it positive}, as might have been expected for real
particles. Thus the stress tensor of the created particles is completely unlike that in the `eternal'  expanding de Sitter background,
where the stress tensor tends to the de Sitter invariant Bunch-Davies attractor value with $\varepsilon + p=0$, all initial state 
deviations from this value falling exponentially with time \cite{Attract}. This occurs because the oscillatory phase coherent terms 
do {\it not} wash out at late times in fixed de Sitter space, as they do in Fig.~\ref{Fig:posc}, but instead give a contribution of the 
same order as that of the created particles, combining with them to give the de Sitter invariant value at late times. 
This phase coherence is due to the fact that all the Fourier modes in the broad range of values $\g \ll k \lesssim \g a(t)$ remain in phase,
because of the exponential suppression of both the $t$ and $k$ dependence of (\ref{ThetadS}), through $k/a = k e^{-Ht}$ in 
de Sitter space. Thus as these modes pass outside the de Sitter Hubble horizon, they have nearly the same time dependence 
and add coherently in the integral over $k$, remaining of the same order as the particle creation terms. Our results show that this
phase coherence of superhorizon modes in pure de Sitter space is destroyed by the transition out of de Sitter, however gentle, while the 
particle number term $|B_k|^2$ is robust, surviving the transition due to its adiabatic invariance.

In order to estimate the backreaction of the created particles, we note that the Einstein equation
\be
\frac{dH}{dt} = - \frac{4 \pi G}{3}\,  (\varepsilon + p) < 0
\label{Hless}
\ee
for $\varepsilon + p >0$ in the final state, tending to decrease the curvature, assuming that the phase coherence of the superhorizon 
modes is not preserved and the particle contributions dominate the stress tensor finally. From Eqs.~(\ref{enerpress}) we have
\be
\varepsilon + p = \frac{H^4\,|B_{\g}|^2 \g^3\!}{6 \pi^2}\ \ka^3\left(\ka^2\g^2 +\sdfrac{m^2}{H^2}\right)^{\!\frac{1}{2}}> 0 
\ee
so that (\ref{Hless}) leads to a fractional decrease in the expansion rate of order
\be
\frac{\D H}{H} \simeq -\frac{2}{9 \pi}\,GH^2\, \frac{\g^4 \,\ka^3}{e^{2\pi\g}-1}\left(\ka^2 + 1+ \sdfrac{1}{4\g^2}\right)^{\!\frac{1}{2}}
\label{DelH}
\ee
for a Hubble expansion time $H \D t \simeq1$. The backreaction is small if $GH^2 \ll 1$ and contains the additional exponential suppression from 
$|B_{\g}|^2$ as in (\ref{RateBDlargem}) if $m \gg H$. It is nevertheless non-zero and appears at one-loop order even for a massive free field, 
in contrast to quantized graviton contributions reported at two-loop order in \cite{Wood}. 

Note that although the energy density and pressure (\ref{enerpress}) are not exponentially redshifted away due to the constant rate 
of particle creation in de Sitter space, neither do they grow in the time $T$ that the de Sitter phase persists, as the electric current does 
in the $E$-field case, {\it c.f.} Fig.~\ref{Fig:jzoscT}. In the $E$-field case the created particles are accelerated to relativistic velocities 
after their creation and contribute a current which grows linearly with the window of modes that go through their creation events, and 
hence that grows secularly with time, producing a backreaction effect on the electric field that clearly must eventually be taken into 
account in a consistent dynamical system. This acceleration of created particles to relativistic velocities appears more
similar to the contracting part of the time slicing of the full de Sitter hyperboloid, in which the created particles are blueshifted rather
than redshifted and exponentially growing stress-energy perturbations occur \cite{AndMot1,AndMot2}.

\section{Discussion}
\label{Sec:Concl}

In this paper we have presented a detailed analysis of particle creation and vacuum decay in persistent background
fields that are homogeneous in space, such as the constant uniform electric field and de Sitter space. The vacuum state of QFT
in the presence of such background external fields is specified not by analytic continuation to Euclidean time, but by the Feynman-Schwinger
$m^2-i\e$ causal prescription which covariantly defines particle and antiparticle excitations, and their absence, in real time.
This defines a scattering problem (\ref{asymBog}) for massive scalar fields which determines the mean particle number created 
in pairs in each Fourier mode, and relates the vacuum persistence probability (\ref{inout}) directly to the number of created particles. 
Let us emphasize that the zero overlap between the $| in\rag$ and $| out\rag$ states in (\ref{inoutrate}) in the strict limit of infinite 
four-volume is no pathology as is sometimes implied \cite{BirDav}, but simply a consequence of a constant vacuum decay rate $\G$ 
per unit time per unit volume in a persistent background field. The four-volume ${\cal V}_4$ must be extracted from a proper order of limits, 
as by turning the persistent background on and off again after a long time $T$ before evaluating the Integral Rate (\ref{DecayRateInt}).

By analyzing the particle creation process in real time, we have also given an invariant Differential Rate formula (\ref{DecayRateDiff}) for 
the vacuum decay rate in such persistent fields in which no divergent integrals over momenta are encountered. The evaluation of this 
Differential Rate relies upon an analysis of the critical points of the adiabatic phase integral (\ref{genTheta}) in the complex time domain, 
and the semi-classical definition of the time at which this particle creation can be said to occur. This time $t_{event}({\bf k})$ is defined by the point 
at which the Stokes line of constant Real Part of the adiabatic phase for the given Fourier mode $\bf k$ crosses the real time axis, and thereby 
gives a relation between $\bf k$ and $t$ that determines the Jacobian in (\ref{DecayRateDiff}). In the case of the constant, uniform electric field, 
Schwinger's result for the decay of the vacuum into charged particle/antiparticle pairs is recovered in this way. In the case of de Sitter space, 
the pre-factor of the Bunch-Davies vacuum decay rate, undetermined by earlier treatments, is also fixed, with the principal result being (\ref{DecayBD}).

We have also discussed an Integral Method (\ref{DecayRateInt}) for calculating the vacuum decay rate in persistent background
fields, which relies upon replacing the external field extending infinitely to the past and future in time, by one which is
adiabatically switched on from zero around some finite time $t_0$, allowed to persist for a long but finite time until $t_1$,
and then adiabatically turned back to zero again. This defines the total number of particles in the asymptotic final state unambiguously
after the background electric or gravitational field is turned off, and verifies that the adiabatic particle number definition ${\cal N}_k^{(n)}$ for 
either $n=1,2$ is robust, giving the correct average number of asymptotic particles in a given Fourier mode. For this Integral Method
of defining the vacuum decay rate of a persistent field to work, it is necessary to find a time dependent background for which  any effects associated 
with switching the background field on and off can be made negligibly small in the limit $T= t_1 -t_0 \rightarrow \infty$. We found a suitable 
two-parameter family of external gauge potentials (\ref{Eprofile}) for which this condition is satisfied, and once again found the Schwinger 
decay rate for a long-persistent uniform electric field by this Integral Method.

In the case of de Sitter space, the apparently natural generalization of this two-parameter FLRW background spacetime (\ref{a2param})
does {\it not} yield the de Sitter decay rate, because the four-volume factor is multiplicative rather than additive in time, and the
asymptotic particle number depends upon the time scale $b^{-1}$ with which the de Sitter background is turned off at late 
times, no matter how long the de Sitter phase lasts. Since the two-parameter FLRW background (\ref{a2param}) is the switching off of 
de Sitter background curvature everywhere in space in cosmic time, including outside the de Sitter-Hubble horizon, one might suspect that 
this spatially homogeneous background is particularly artificial, and perhaps should be replaced with one that is regulated also in its spatial 
extent at the horizon scale. The failure of the Integral Method for strictly spatial homogeneous switching on/off of de Sitter space seems to be 
indicative of a greater sensitivity of de Sitter space to spatial boundary conditions than in the electric field case, due to the effect 
of long wavelength modes lying outside their causal Hubble horizon. 

The main conclusion to be drawn from the existence of particle creation and a non-zero decay rate (\ref{DecayBD}) starting from
the de Sitter invariant Bunch-Davies state is that this CTBD state is not a stable ground state of QFT in de Sitter space, and that
$SO(4,1)$ de Sitter symmetry is necessarily broken, both in time, and possibly also in space, even by a free massive quantum field 
without self-interactions. Stated differently, the Feynman-Schwinger $m^2 - i 0^+$ definition of the vacuum of QFT, its particle excitations 
and its propagator function is incompatible with the requirements of de Sitter invariance and Euclidean ${\mathbb S}^4$ boundary
conditions on propagators, at least for conformally coupled massive scalar fields with any finite $m > H/2$, for which the
particle concept is well-defined.

This incompatibility of Euclidean continuation and the specification of the vacuum in real time follows from the fact that justification for the
analytic continuation to Euclidean time relies upon the system possessing a Hamiltonian $\bf \hat H$ that is both time-independent and 
bounded from below, guaranteeing that a stable vacuum exists. The example of the constant, uniform electric field which does not possess
such a Hamiltonian shows that continuation to Euclidean time is inconsistent with particle creation and vacuum decay expected in such a 
background $E$-field. This example of spontaneous vacuum decay shows that the $m^2-i\e$ definition of particle excitations 
spontaneously breaks the time reversal symmetry of the background \cite{Fluc}, although an otherwise Lorentz invariant vacuum decay 
rate $\G = 2\, {\rm \cL}_{\rm eff}$ can be defined. 

At the particle worldline level the vacuum decay is typified by hyperbolic trajectories of constant acceleration of  particles in an electric field, in 
contrast to the closed circular orbits in the corresponding Euclidean constant magnetic field, which does possess a stable QFT vacuum. The 
hyperbolic trajectories of freely falling test particles in de Sitter space, drawn away from each other by the de Sitter expansion are similarly 
clearly different qualitatively from the closed circular trajectories of test particles on the compact Euclidean ${\mathbb S}^4$ manifold. Since 
analytic continuation of propagators to the Euclidean ${\mathbb S}^4$ manifold enforces boundary conditions whose semi-classical limit 
are precisely these closed circular trajectories, it defines a theory inequivalent to that on the Lorentzian de Sitter manifold which requires 
quite different boundary conditions at asymptotic early and late times $t \rightarrow \mp\infty$, even in the absence of self-interactions. 
These different boundary conditions lead to the instability of the Bunch-Davies state to particle creation in real time.

Since the Feynman-Schwinger $m^2 - i \e$ condition goes smoothly over to that of the standard Minkowski vacuum for any slowly time-varying adiabatic 
background, whether electromagnetic or gravitational, independent of any symmetry of the background, analyticity in the mass parameter is a more 
general principle of determination of the vacuum of QFT, more firmly based on physical considerations of causality than the Euclidean postulate.
The compatibility of Wick rotation in time to the $m^2 \rightarrow m^2 - i \e$ prescription is a special property of zero-field Minkowski space
where Poincar\'e invariance dictates that correlation functions at $x$ and $x'$ can only be a function of $m^2(x-x')^2 = -m^2 (t-t')^2 + m^2 (\bx -\bx')^2$,
with the result that analyticity in $m^2$ and $t$ are necessarily related. This equivalence cannot be assumed in general, and in particular it ceases
to hold when additional parameters of the background field enter the time dependence of correlation functions, when there is no invariant decomposition
into positive and negative frequency subspaces, or when a Hamiltonian bounded from below does not exist in real time, as in de Sitter space, in which
cases continuation to Euclidean time has no evident physical justification.

On any given FLRW time slice of constant  $t$ in the spatially flat coordinates (\ref{dSflat}), the physical adiabatic vacuum is Bunch-Davies 
only for Fourier modes with wave numbers $k \gg \overline k(t) = \ka\g e^{Ht}$, while for modes with $k \ll \overline k(t)$ the vacuum state is 
described by the positive frequency {\it out} mode functions (\ref{outflat}), with a smooth but fairly rapid switchover at $k \sim \overline k(t)$. 
Since modes continue to redshift with the de Sitter expansion, the dividing line $\overline k(t)$ between the modes in the Bunch-Davies 
vacuum and those whose vacuum state is defined by (\ref{outflat}) continues to grow in co-moving wavenumber $k$. This implies
that particles are continuously created at $k\sim \overline k(t)$, and both the vacuum decay rate and the total energy density and pressure
of the created particles is independent of the duration of the de Sitter expansion.

A by-product of the present study of particle creation is that adiabatic particle number is robust, in the sense that its asymptotic
value in a time-dependent background survives the adiabatic switching off of that background and agrees with the clear and
unambiguous definition of particles in the final zero-field flat space region. The interesting consequence of the kinematics of particle
creation in de Sitter space, and these particles surviving after the de Sitter background has been switched off is that they contribute
a constant and in fact positive energy density and pressure in the final state, which does not redshift away, no matter how long the de Sitter
phase persists. This is in contrast to the situation in an eternally fixed pure de Sitter background, where despite the particle creation
there are in addition phase coherent terms similar to the $A_{\bf k}B^*_{\bf k}$ quantum interference terms in (\ref{jz}) and (\ref{press})
which conspire to exactly cancel the particle creation terms and restore the de Sitter invariant Bunch-Davies value of $p= -\varepsilon$
asymptotically at late times \cite{Attract}. Thus the late time pure de Sitter limit is {\it not} equal to the the late time limit of a FLRW time profile
such as (\ref{a2param}), no matter how large the finite time $t_1$ is taken, and no matter how gently the de Sitter phase ends. 

We have estimated the magnitude of the backreaction effect by (\ref{Hless})-(\ref{DelH}) in the case that exact spatial homogeneity is preserved, 
assuming the quantum phase oscillation terms can be neglected. The effect is small for massive fields if $GH^2 \ll 1$. Nevertheless {\it any} 
instability of de Sitter space due to particle creation effects indicates that it is not the stable ground state of QFT coupled to Einstein gravity with a 
cosmological constant, and these particle creation effects should be taken into account in a fully consistent backreaction {\it in-in} calculation. 
We have not considered light or massless fields in this paper, but one may suspect that their backreaction effects could be significantly larger.
In these cases where $\Theta_k$ becomes pure imaginary on the real time axis, it would be better to consider the modes not as `particles,'
but as coherent waves. If these superhorizon modes of a massless field decohere, the remaining excitations above the vacuum at the end of the 
de Sitter phase could have important consequences for the reheating of the universe at the end of inflation.

In exact eternal de Sitter space the $O(4,1)$ de Sitter invariant Bunch-Davies state is a self-consistent solution of the semi-classical 
Einstein eqs.~for all times. Thus some variation from the Bunch-Davies state is necessary in the initial conditions, which is natural if the 
de Sitter phase begins at a finite time $t_0$ rather than in the infinite past, {\it and} some variation of the background from exact de Sitter
is necessary in order to upset the exact phase coherence of the quantum interference terms of the superhorizon modes in de Sitter space. Only if 
both these elements are present can the quantum interference terms be neglected in a self-consistent backreaction calculation, and the 
particle creation terms with $\varepsilon + p >0$ may slowly but surely reduce the effective cosmological `constant' $\L_{\rm eff} = 3H^2$ according 
to (\ref{Hless}). For such a phase decoherence mechanism to work, the decoherence time scale for the long wavelength modes with $k \lesssim \overline k (t)$ 
must be less than or of the order the Hubble expansion time $H^{-1}$. Although this has been speculated even for free fields \cite{BurHolTasMarkk}, 
decoherence is likely even more effective when backreaction, self-interactions and loop effects are taken into account, where the infrared sensitivity of the
Bunch-Davies correlations becomes more apparent \cite{Poly,AkhBui,Akh,AkhMosPavPop}. We remark that in taking account of interactions, through 
Boltzmann transport equations, the adiabatic particle definition of particles provides the link between QFT and a fully classical (completely 
phase incoherent) particle limit, when the distinctions between different adiabatic orders should also become unimportant. 

Finally the sensitivity of the rate (\ref{Fig:Rateb}) on how the de Sitter phase ends simultaneously over all space at late times in the
time profile (\ref{a2param}) and the sensitivity of the stress tensor to phase coherence/decoherence of the same $k \sim \overline k(t)$ 
modes appear to be related. If the persistent de Sitter background were to be regulated differently, in a way consistent with a finite causal
Hubble horizon, by modifying it with a spatial regulator rather than switching it on and off in FLRW time everywhere in space, these superhorizon 
modes would be treated quite differently, or cut off entirely. Hence it appears likely that sensitivity to spatial boundary conditions through a regulator or
other physics on the horizon scale will survive in a more complete treatment. If so, this would imply spatial homogeneity is broken on 
the horizon scale $H^{-1}$ as well, leading to a spatially inhomogeneous rather than global FLRW cosmology. Evidence for the additional breaking 
of spatial homogeneity, as well as time reversal inherent in vacuum decay of de Sitter space, was presented in the closed spatial sections 
previously in Ref.~\cite{AndMot2}. If spatial homogeneity is broken, the backreaction of the created particles in a spatially inhomogeneous 
universe should be considered, with potentially far-reaching consequences for spacetime dependent vacuum dark energy and observational 
cosmology on the scale of the Hubble horizon, even in the present epoch.

\vspace{-4mm}
\begin{acknowledgments}
\vspace{-4mm}
E. M. acknowledges a stimulating discussion with A. M. Polyakov prior to publication of this work, and useful comments on a first draft by
E. T. Akhmedov, C. Burgess, and S. A. Fulling. This work was supported in part by the National Science Foundation under Grants 
No. PHY-0856050, No. PHY-1308325, and No. PHY-1505875 to Wake Forest University.  Some of the numerical work was done 
using the WFU DEAC cluster; we thank the WFU Provost's Office and Information Systems Department for their generous support. 
Numerical work relating to this project was also done using the OIT henry2 cluster at North Carolina State University.
\end{acknowledgments}

\end{document}